**Universal electrical resistivity in iron-based high-temperature superconductors**

Evgeny F. Talantsev[1,2*]

[1] M.N. Mikheev Institute of Metal Physics, Ural Branch, Russian Academy of Sciences, 18 Sofia Kovalevskaya St, Ekaterinburg 620108 Russia

[2] Nanotech Centre, Ural Federal University, 19 Mira St, Ekaterinburg 620002 Russia

*evgeny.talantsev@imp.uran.ru

## Abstract

Temperature-dependent electrical resistivity $\rho(T)$ in cuprate, iron-based, and nickelate superconductors represents one of unresolved problems in modern solid-state physics, because of the high absolute value of $\rho(T)$, which, in many cases, exceeds the Ioffe-Regel criterion, and a large difference in the $\rho(T)$ shape for samples with different doping. Here I show that the nearest-neighbour hopping parallel resistivity model can well describe the in-plane resistivity $\rho_{ab}(T)$ in iron-based superconductors (IBS) of type 11, 111, 122, 1111, 1144 and 42214 in which the spin density wave phase or nematic order is suppressed by doping. The model is based on the postulate that $\rho_{ab}(T)$ is a sum of two parallel conduction channels, where one channel is described by the Bloch-Grüneisen equation and the other by the Arrhenius hopping equation. Consequently, the variety of $\rho_{ab}(T)$ in the IBS is explained as a variation of the Debye temperature $\Theta_D$ and the hopping activation energy $E_a$. The derived Debye temperatures were used to estimate the electron-phonon coupling constant in the studied IBS.

## I. Introduction

Iron-based superconductors (IBS) discovered by Hosono's group two decades ago [1] represent perhaps the richest family of high-temperature superconductors (HTS) in phases, chemical and structural composition [2–17], among other families of HTS [18–23]. The physical, chemical, and superconducting properties, phase transformations, spin density correlations in IBS and related compounds have been extensively studied since then [15,24–64]. There are several ongoing projects for the development of manufacturing technology for the practical use of IBS [65–72].

Temperature-dependent electrical resistivity $\rho(T)$ represents one of the most common measurements in the discovery and study of superconductors [73–98]. There are two commonly accepted approaches to analyzing $\rho(T)$ data measured in the IBS, one of which is the second-order polynomial fitting [99–104]:

$$\rho(T) = \sum_{i=1}^{N} \rho_i(T) = \sum_{i=1}^{N} A_i \times T^i \quad (1)$$

where, $i = 0,1,2$, and $A_i$ are free-fitting parameters. This approach has three main problems. First, it works only in a narrow temperature range well above $T_c$ [105]. Second, the obtained parameters are highly dependent on the temperature range chosen for fitting. Third, the obtained parameters $A_i$ can change significantly (or even change sign) with a small change in the experimental $\rho(T)$ curve.

The second common approach is the power-law fit [100,101,106,107]:

$$\rho(T) = \rho_0 + A \times T^{\mathrm{n}}, \quad (2)$$

where $\rho_0$, $A$, and n are free-fitting parameters. First, it also works only in a narrow temperature range well above $T_c$ [107]. Another problem with this approach is that the deduced parameters $A$ have different physical units $[\Omega\mathrm{mK}^{-\mathrm{n}}]$, and therefore the interpretation and comparison of the deduced parameters for different samples are impossible. Other problems with this approach were discussed in Ref. [108].

In overall, Equations 1,2 are based on Matthiessen's rule [109–113]:

$$\rho(T) = \sum_{i=1}^{N} \rho_i(T), \tag{3}$$

where $\rho_i(T)$ represents a direct summation contribution to the resistivity from the $i$-th dissipation mechanism. It should be noted, that the Matthiessen's rule relies on very little physical basis for single-phase metallic conductors (in particular for the in-plane resistivity of single crystals, $\rho_{\mathrm{ab}}(T)$), because electric current flows through the entire cross-section of conductors in the normal state (this differs from the non-dissipative transport current flow in superconductors [63–66]), and, consequently, the contribution of each dissipation mechanism should be a sum of their conductivities:

$$\rho(T) = \frac{1}{\sum_{i=1}^{N} \frac{1}{\rho_i(T)}}, \tag{4}$$

because each dissipation mechanism contributes as a resistance in the parallel electrical circuit.

This fundamental paradigm was taken into account in the parallel resistivity model [114–116]:

$$\rho(T) = \frac{1}{\frac{1}{\rho_{sat}} + \frac{1}{\rho_{BG}}} = \frac{1}{\frac{1}{\rho_{sat}} + \frac{1}{\rho_0 + \rho_D \times \left(\frac{T}{\Theta_{\mathrm{D}}}\right)^p \times \int_0^{\frac{\Theta_{\mathrm{D}}}{T}} \frac{x^p}{(e^x - 1)(1 - e^{-x})} dx}}, \tag{5}$$

where $\rho_{sat}$, $\rho_0$, $\rho_D$, $\Theta_{\mathrm{D}}$ are free fitting parameters, $p = 5$ for pure electron-phonon scattering mechanism (if the dominant dissipation mechanism is different, then $p$ exhibits different values [117]); and $\rho_{sat}$ is temperature-independent free-fitting parameter~~s~~, which is related [114] to the Ioffe-~~Rigel~~ Regel [118] saturated resistivity, $\rho_0$ is the residual resistivity, $\rho_D$ is the resistivity constant, and $\Theta_{\mathrm{D}}$ is the Debye temperature.

This model (Equation 5) is used to deduce the Debye temperature $\Theta_{\mathrm{D}}$ in a wide range of low-temperature superconductors and hydride superconductors [119–126]. As it was clarified by Nowak et al. [127], $\Theta_{\mathrm{D}}$ deduced by Equation 5 should be designated as $\Theta_{\mathrm{D}}^{\mathrm{HT}}$ because it describes the phonon spectrum at high temperature range, above the transition temperature $T \gtrsim 1.5 \times T_{\mathrm{c}}$. In addition to that, in many studies, the low-temperature Debye temperature $\Theta_{\mathrm{D}}^{\mathrm{LT}}$ is deduced from the low-temperature heat capacity data $C_{\mathrm{p}}(T)$ [121,123,126–130]:

$$C_{\mathrm{p}}(T) = \gamma_n \times T + \beta_3 \times T^3 + a_5 \times T^5, \qquad (6)$$

where $\gamma_n$ is the Sommerfeld coefficient, $\beta_3$ is the lattice specific heat coefficient, and $a_5$ is the correction term for phonon dispersion and anharmonicity. From the deduced $\beta_3$ value, the low-temperature Debye temperature is calculated via the formula [127]:

$$\Theta_{\mathrm{D}}^{\mathrm{LT}} = \left(\frac{12\pi^4 \times N \times R}{5 \times \beta_3}\right)^{\frac{1}{3}}, \qquad (7)$$

where $N$ is the number of atoms per formula unit (f.u.), and $R = 8.31446\ [\mathrm{J\ mol^{-1}K^{-1}}]$ is the universal gas constant.

There are always some differences between $\Theta_{\mathrm{D}}^{\mathrm{LT}}$ and $\Theta_{\mathrm{D}}^{\mathrm{HT}}$ because these two values are characteristic temperatures for different parts of the phonon spectrum.

It should be noted that (to the best of my knowledge) all reported in the literature values of the Debye temperature $\Theta_{\mathrm{D}}$ for iron-based superconductors represent the values of the Debye temperature at low temperatures $\Theta_{\mathrm{D}}^{\mathrm{LT}}$ [131,132]. In contrast, in this study, since the analysis is based on $\rho(T)$ data equal to zero at $T < T_c$, the derived Debye temperature values $\Theta_{\mathrm{D}}$ reported below must be assigned to $\Theta_{\mathrm{D}}^{\mathrm{HT}}$.

There are very rare studies [133,134] where Equation 5 is applied for unconventional HTS. Some research groups utilize different reduced forms of Equation 5 [135]:

$$\rho(T) = \frac{1}{\frac{1}{\rho_{sat}}+\frac{1}{\rho_0+A\times T}}, \quad (8)$$

or [136]:

$$\rho(T) = \frac{1}{\frac{1}{\rho_{sat}}+\frac{1}{\rho_0+A\times T^2}}, \quad (9)$$

or [137]

$$\rho(T) = \frac{1}{\frac{1}{\rho_{sat}}+\frac{1}{\rho_0+A\times T+B\times T^2}}, \quad (10)$$

where $\rho_{sat}$, $\rho_0$, $A$, and B are free-fitting parameters.

Shoemaker et al. [138] to approximate in-plane temperature-dependent resistivity $\rho_{\mathrm{ab}}(T)$ in single crystal $K_{0.85}Fe_{1.9}As_2$ proposed the following equation:

$$\rho(T) = \frac{1}{\frac{1}{\rho_{metal}(T)}+\frac{1}{\rho_{semiconductor}(T)}} = \frac{1}{\frac{1}{\rho_0+A\times T^{\mathrm{n}}}+\frac{1}{\rho_{\mathrm{sat}}\times e^{\frac{E_{\mathrm{g}}}{2\mathrm{k_B}T}}}}, \quad (11)$$

where $\rho_0$, $A$, n, $\rho_{\mathrm{sat}}$, and $E_{\mathrm{g}}$ are free fitting parameters, $E_{\mathrm{g}}$ is insulator activation energy, and $\mathrm{k_B} = 1.3806 \times 10^{-23}$ [$\mathrm{JK^{-1}}$] $= 8.6173 \times 10^{-5}$ [$\mathrm{eVK^{-1}}$] is the Boltzmann constant.

It should be noted that Equation 11 was used to analyze the $\rho_{ab}(T)$ data measured in other single crystals of iron chalcogenides [139,140].

The left side of Equation 11 is the exact main postulate on which a universal temperature-dependent model for the electrical resistivity of pure actinide metals has recently and independently been proposed [141]. Now, I must acknowledge that the postulate and primary idea for the model [141], which was named the nearest-neighbour hopping parallel resistivity model, had been proposed by Shoemaker et al. [138].

Since it is difficult to assume that pure elemental actinide metals can exhibit a semiconducting band, in [141], instead of $\rho_{semiconductor}(T)$ in the left-hand side of Equation 11, I used a more general assumption, according to which the temperature-independent saturation resistivity in the parallel resistivity model [114–116], $\rho_{sat}$, exhibits a temperature dependence that mathematically obeys the Arrhenius-Mott-Efros-Shklovskii type of law [142,143], however, the specific physical mechanism for which has not yet been established (it can be spin fluctuations):

$$\frac{1}{\rho_{sat}} \rightarrow \frac{1}{\rho_{sat} \times e^{\left(\frac{E_a}{k_B T}\right)^m}}, \tag{12}$$

where $E_a$ is the activation energy for the hopping, and $m$ is determined by the particular hopping mechanism. Also, in Ref. [141] exact formula for the $\rho_{metal}(T)$ part was used:

$$\frac{1}{\rho_{metal}(T)} \rightarrow \frac{1}{\rho_0 + \rho_D \times \left(\frac{T}{\Theta_D}\right)^{p=5} \times \int_0^{\frac{\Theta_D}{T}} \frac{x^{p=5}}{(e^x - 1)(1 - e^{-x})} dx}. \tag{13}$$

The analysis of the $\rho(T)$ of pure actinide metals [141] showed that $m$ in Equation 12 can be taken equal to 1, and, thus, the analysis was performed by the following equation:

$$\rho(T) = \frac{1}{\frac{1}{\rho_{sat} \times e^{\frac{E_a}{k_B T}}} + \frac{1}{\rho_0 + \rho_D \times \left(\frac{T}{\Theta_D}\right)^{p=5} \times \int_0^{\frac{\Theta_D}{T}} \frac{x^{p=5}}{(e^x - 1)(1 - e^{-x})} dx}}. \tag{14}$$

As mentioned in [141], the condition $p = 5$ in Equation 14 means that the charge carriers exhibit an electron-phonon dissipation mechanism; if the dominant mechanism is different, $p$ should have a different value [108,144].

Equation 14 allows to determine the Debye temperature $\Theta_D^{HT}$ besides the activation energy $E_a$, while interpretations of the power-law exponent $n$ and parameter $A$ in Equation 11 are unknown.

It is important to note that Equations 11 and 14 have the same number of free fitting parameters, five. However, Equation 14 approximates the experimental $\rho_{ab}(T)$ data with much higher accuracy, which can be seen in Figure 1, where the $\rho_{ab}(T)$ data reported by Roslova et al. [140] for single crystal of iron chalcogenide $(K_{1-x}Na_x)_yFe_{1.75}Se_2$ (x=0.32; y = 0.95) are fitted with two Equations.

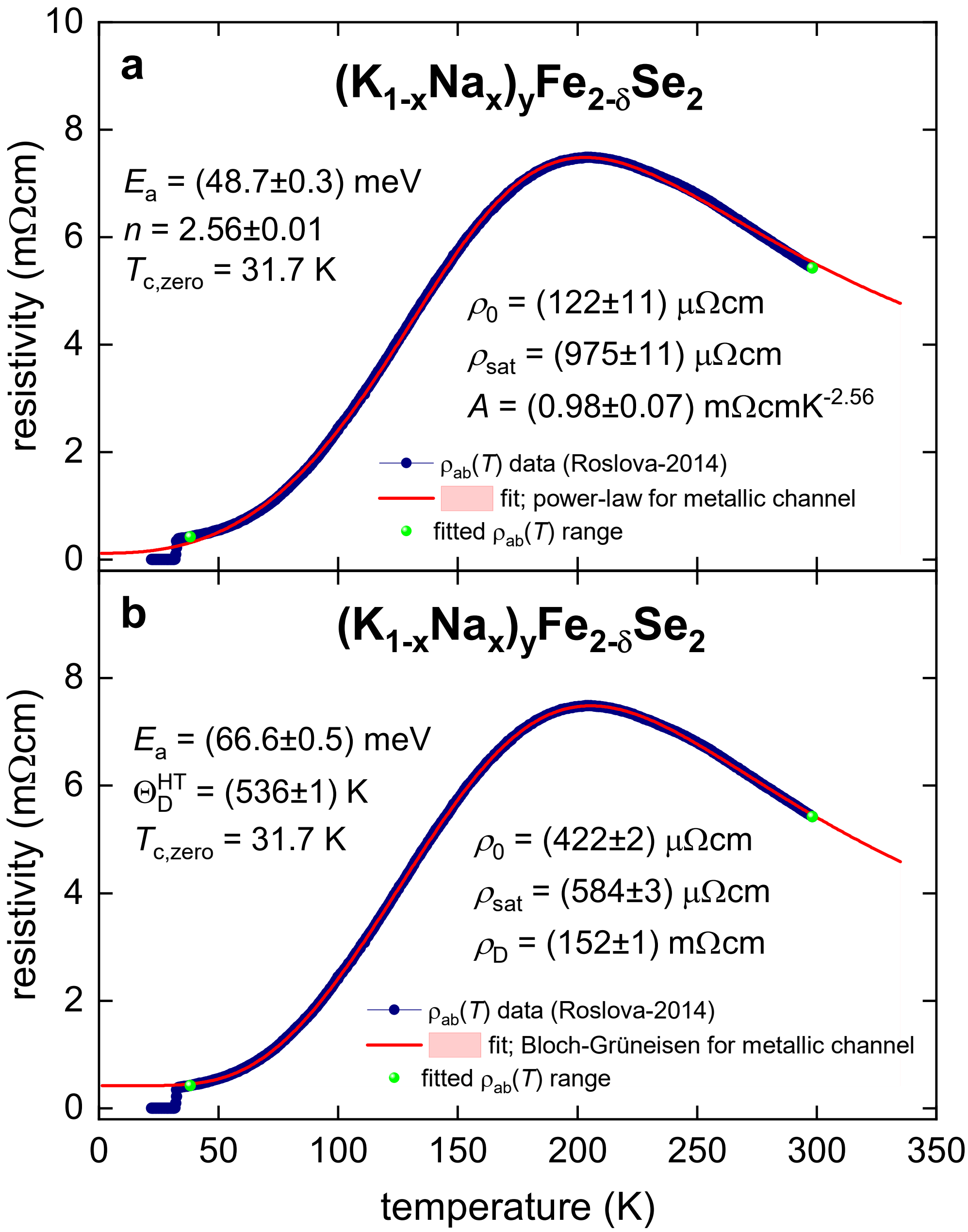

**Figure 1.** Fits with (**a**) Equation 11 [138], and (**b**) Equation 14 [141] of the same $\rho_{ab}(T)$ data measured in 122-type chalcogenide single crystal $(K_{1-x}Na_x)_yFe_{1.75}Se_2$ (x=0.32; y = 0.95) (data reported by Roslova et al. [140]). (**a**) Equation of $\frac{E_g}{2} = E_a$ was used to calculate $E_a$. Fit quality $R$-square (COD) = 0.9997. (**b**) Fit quality $R$ = 0.9999. Deduced parameters are shown in each panel. The 95% confidence bands are shown by pink shadow areas.

The analysis of in-plane temperature-dependent resistivity $\rho_{ab}(T)$ in iron-based superconductors (presented below) has shown that Equation 14 perfectly approximates the experimental data and that the deduced parameters are within the approximate range reported by independent experimental studies (in particular, the Debye temperature values).

Since the specific hopping mechanism for the second conduction channel is unknown, I continue to use the designation $E_{\mathrm{a}}$ in Equation 14. However, if it were proven that the mechanism belongs to a pure semiconductor type (as it was assumed by Shoemaker et al. [138]), the deduced activation energy $E_{\mathrm{a}}$ can be recalculated to the semiconducting gap energy $E_g$:

$$E_g = 2 \times E_{\mathrm{a}} \quad (14)$$

It is also worth mentioning that the analysis presented here was carried out for high-temperature iron-based superconductors in which the formation of the spin-density wave phase is suppressed by chemical doping.

It should also be noted that since the IBS family is the largest among high-temperature superconductors (HTS), it is practically impossible to present results for all types of chemical doping or compressed IBS in a single paper. Therefore, the analysis is presented for typical chemical doping to demonstrate the universal applicability of the nearest-neighbours hopping parallel resistivity model (Equation 14) for analysing the in-plane electrical resistivity $\rho_{ab}(T)$ measured in single crystals of IBS phases 11, 111, 122, 1111, 1144, and 42214.

## II. Data and Software

Because all IBSs are layered quasi-2D superconductors, thermodynamic fluctuations of the order parameter are significant in these superconductors [104,145,146]. Based on this, an excess in electrical conductivity appears in quasi-2D superconductors well above the transition temperature $T_c$ [104,147–152]. Since in this study the analysed experimental data are the temperature-dependent electrical resistivity $\rho_{ab}(T)$, to avoid access to conductivity due to thermodynamic fluctuations of the superconducting order parameter, the experimental data $\rho_{ab}(T)$ were fitted with Equation 14 for the temperature range $T \gtrsim 1.5 \times T_{c,zero}$.

All analysed $\rho_{ab}(T)$ datasets were obtained by digitizing original experimental studies by using the *ScanIt* software [153], version 2.0.8.0. (Programming: Jasper van Baten, thanks to: Richard Baur). Data fits were performed in the *Origin* software [154], version 2021.

## III. Results

### 3.1. Single crystals of the 1111-type

Iron-containing superconductors were discovered by Kamihara et al [1] in 2006 by observing superconductivity in $LaFePO_4$ [1]. The highest superconducting transition temperature is also observed for compounds in this 1111-type [24,71]. It is therefore natural to begin by presenting the results of the analysis obtained for compounds of this 1111-type.

Ma et al. [155] grew $Ca(Fe_{1-x}Co_x)AsF$ single crystals using the self-flux method and reported $\rho_{ab}(T)$ data for $Ca(Fe_{1-x}Co_x)AsF$ (x = 0.118) crystal, which has a zero resistance transition temperature $T_{c,zero} = 19$ K [56,155]. The $\rho_{ab}(T)$ data for this single crystal are shown in Figure 2,a along with the fit using Equation 14. The deduced Debye temperature for this sample is relatively high $\Theta_D^{HT} = (599 \pm 6)$ K.

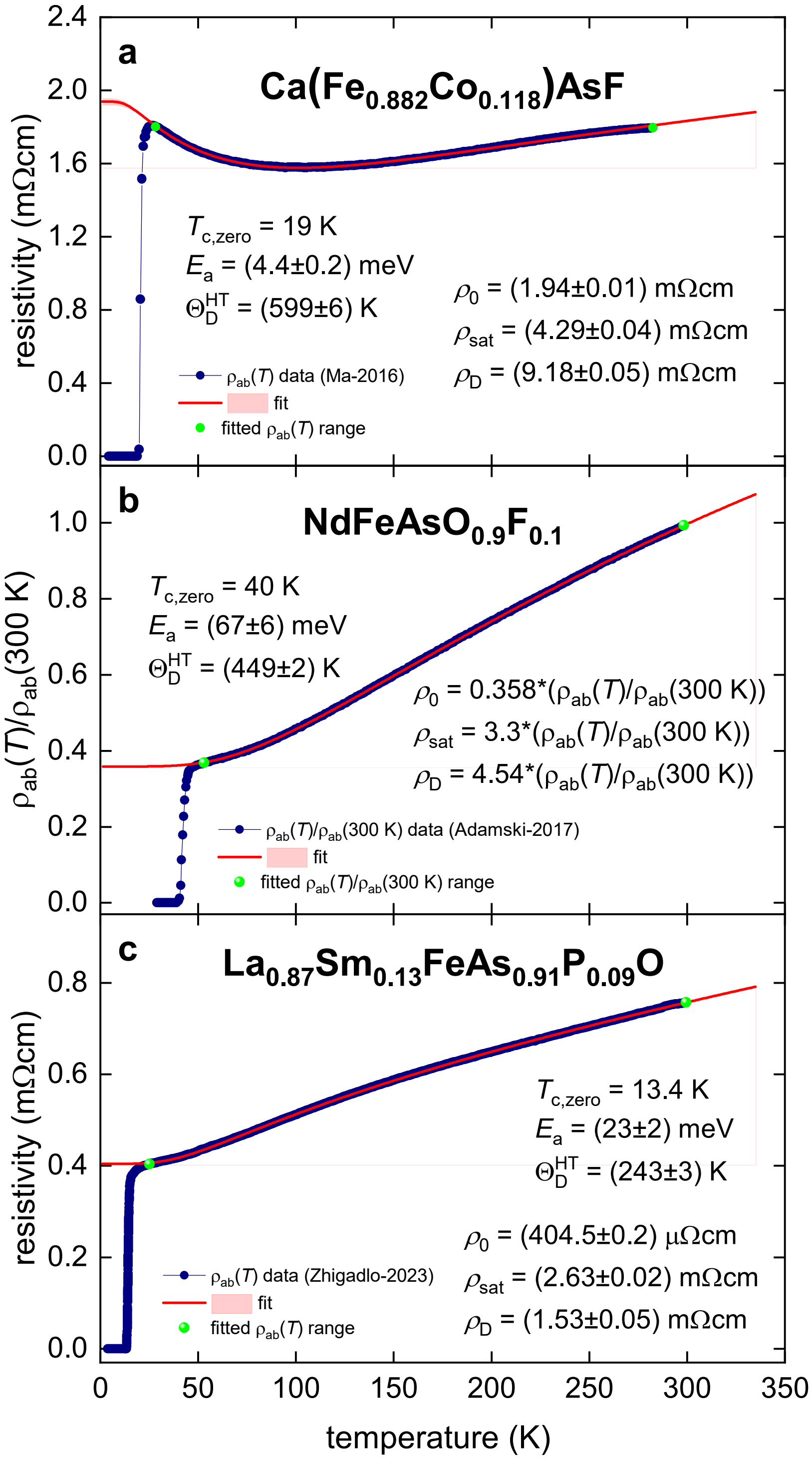

**Figure 2.** Fits with the nearest-neighbor hopping parallel resistance model (Equation 14) of the $\rho_{ab}(T)$ data measured in 1111-type single crystals of the IBS. (**a**) $Ca(Fe_{1-x}Co_x)AsF$ (x=0.118) (data reported by Ma et al. [155]). Fit quality *R*-square (COD) = 0.9979. (**b**) $NdFeAsO_{1-x}F_x$ (x=0.1) (data reported by Adamski et al. [53]). Fit quality *R* = 0.9999. (**c**) $La_{1-x}Sm_xFeAs_{1-y}P_yO$ (x = 0.13, y = 0.09) (data reported by Zhigadlo et al. [156]). Fit quality *R* = 0.9999. Deduced parameters are shown in each panel. The 95% confidence bands are shown by pink shadow areas.

Adamski et al. [53] grew NdFeAsO and $NdFeAsO_{1-x}F_x$ single crystals in the NaCl/KCl flux at ambient pressure and reported the reduced temperature dependence of the resistivity $\frac{\rho_{ab}(T)}{\rho_{ab}(300\text{ K})}$ for three samples. The $\frac{\rho_{ab}(T)}{\rho_{ab}(300\text{ K})}$ data for $NdFeAsO_{1-x}F_x$ (x=0.1) single crystal (which has a zero resistance transition temperature $T_{c,zero} = 40$ K [53]) are shown in Figure 2,b along with the fit using Equation 14. The deduced Debye temperature for this sample is also relatively high $\Theta_{\mathrm{D}}^{\mathrm{HT}} = (449 \pm 2)$ K.

Zhigadlo et al. [156] grew single crystals of $PrFeAsO_{1-y}$ using the cubic-anvil high-pressure and high-temperature technique, and reported $\rho_{ab}(T)$ data for the $La_{0.87}Sm_{0.13}FeAs_{0.91}P_{0.09}O$ single crystal. The $\rho_{ab}(T)$ data for the crystal, which has a zero-resistance transition temperature $T_{c,zero} = 13.4$ K [156]) are shown in Figure 2,c along with the fit using Equation 14. The deduced Debye temperature for this sample is relatively low $\Theta_{\mathrm{D}}^{\mathrm{HT}} = (243 \pm 3)$ K.

However, it should be noted that this sample also has a relatively low superconducting transition temperature $T_{c,zero} = 13.4$ K [156].

The derived primary parameters, $\Theta_{\mathrm{D}}^{\mathrm{HT}}$ and $E_{\mathrm{a}}$, as well as the experimental superconducting transition temperature $T_{\mathrm{c}}$, vary considerably for samples doped at different chemical sites. A detailed analysis of the derived parameters is presented in the *Discussion* section.

### 3.2. Pnictide single crystals of the 122-type

Rotter et al. [6,7] discovered high-temperature superconductivity in the $(Ba_{1-x}K_x)Fe_2As_2$ pnictide 122-type in 2008.

To demonstrate that the nearest-neighbours hopping parallel resistivity model can be used to deduce parameters for non-superconducting pnictide 122-type, in Figure 3,a I show the $\rho_{ab}(T)$ data for single crystal $Ba(Fe_{0.95}Cu_{0.05})_2As_2$ reported by Rosa et al. [157] along with the fit using Equation 14. The single crystal was grown by the in-flux technique [158]. The deduced Debye temperature for this sample is $\Theta_D^{HT} = (420 \pm 3)$ K.

Luo et al. [159] grew $(Ba_{0.6}K_{0.4})Fe_2As_2$ single crystals by flux method using FeAs as flux. The $\rho_{ab}(T)$ data, along with the fit using Equation 14, for the crystal with zero resistance transition temperature $T_{c,zero} = 36$ K [159], are shown in Figure 3,b. The deduced Debye temperature for this sample is $\Theta_D^{HT} = (306 \pm 1)$ K.

Remarkably, the deduced Debye temperature for the single crystal with the same stoichiometry $(Ba_{0.6}K_{0.4})Fe_2As_2$, grown and measured by an independent research group (Chen et al. [159] with $T_{c,zero} = 37$ K), is the same: $\Theta_D^{HT} = (305 \pm 7)$ K (Figure 3,c). It is further worth noting that other parameters deduced for two crystals of $(Ba_{0.6}K_{0.4})Fe_2As_2$ (for instance, $\rho_0$ and $\rho_D$) are also within the same approximate ranges (Figure 2,b,c,).

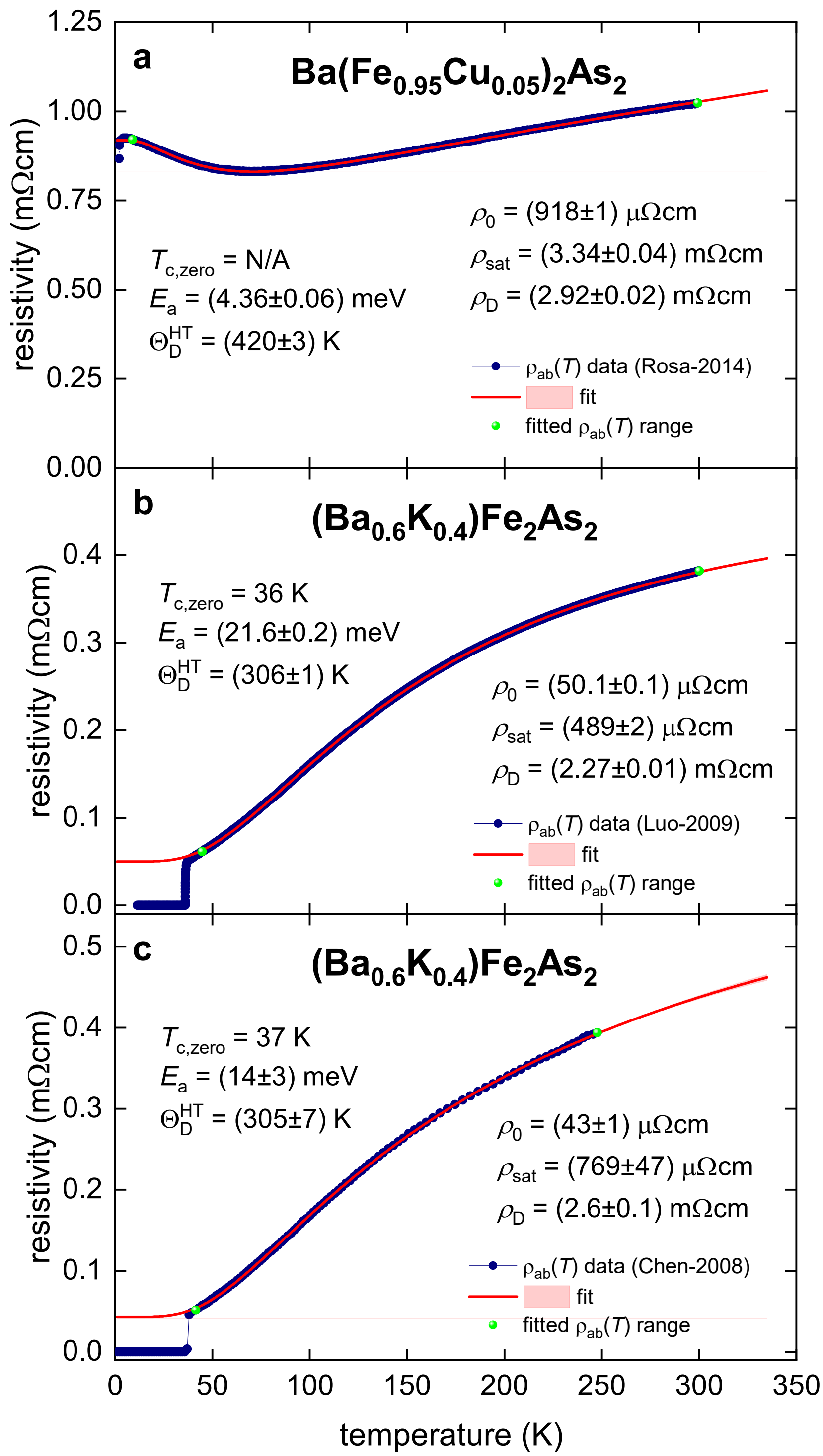

**Figure 3.** Fits with the nearest-neighbor hopping parallel resistance model (Equation 14) of the $\rho_{ab}(T)$ data measured in 122-type pnictide single crystals. (**a**) $Ba(Fe_{1-x}Cu_x)_2As_2$ (x = 0.05) (data reported by Rosa et al. [157]). Fit quality *R*-square (COD) = 0.9996. (**b**) $(Ba_{1-x}K_x)Fe_2As_2$ (x = 0.4) (data reported by Luo et al. [159]). Fit quality *R* = 0.9999. (**c**) $(Ba_{1-x}K_x)Fe_2As_2$ (x = 0.4) (data reported by Chen et al. [160]). Fit quality *R* = 0.9999. Deduced parameters are shown in each panel. The 95% confidence bands are shown by pink shadow areas.

### 3.3. Chalcogenide single crystals of the 122-type

Guo et al. [5] discovered high-temperature superconductivity in the 122-chalcogenide-type single crystals $K_xFe_2Se_2$ in 2010. This is a very rare study in the field of superconductivity, where high-temperature superconductivity was discovered through measurements performed on single crystals.

The first reported $\rho_{ab}(T)$ curve [6] for $K_{0.78}Fe_{1.70}Se_2$ is shown in Figure 4,a along with the fit performed using Equation 14. The deduced Debye temperature for this sample is $\Theta_D^{HT} = (323 \pm 2)$ K and the activation energy for the nearest-neighbours hopping conductivity $E_a = (43.4 \pm 0.1)$ meV. These values agree very well with the values deduced from the experimental curves $\rho_{ab}(T)$ measured in the single crystal $K_{0.85}Fe_{1.90}Se_2$ studied by Shoemaker et al. [138] ($\Theta_D^{HT} = (368 \pm 6)$ K and $E_a = (43.8 \pm 0.1)$ meV (Figure 4,b)) and the single crystal $K_{0.80}Fe_{1.76}Se_2$ studied by Hu et al. [161] ($\Theta_D^{HT} = (357 \pm 1)$ K and $E_a = (38.9 \pm 0.1)$ meV (Figure 4,c)). It should be pointed out that the absolute values of the $\rho_{ab}(T)$ reported by Shoemaker et al. [138] and Hu et al. [161] are different for more than one order of magnitude.

It should also be noted that the deduced:

$$E_a = (43.8 \pm 0.1) \text{ meV} \quad (15)$$

for the $K_{0.85}Fe_{1.90}Se_2$ [138] also agrees well with the reported $E_g$ value by Shoemaker et al. [138], who fit the $\rho_{ab}(T)$ data with Equation 11:

$$E_a = \frac{E_g}{2} = 41.5 \text{ meV} \quad (16)$$

In more details the deduced parameters are discussed in the Discussion section.

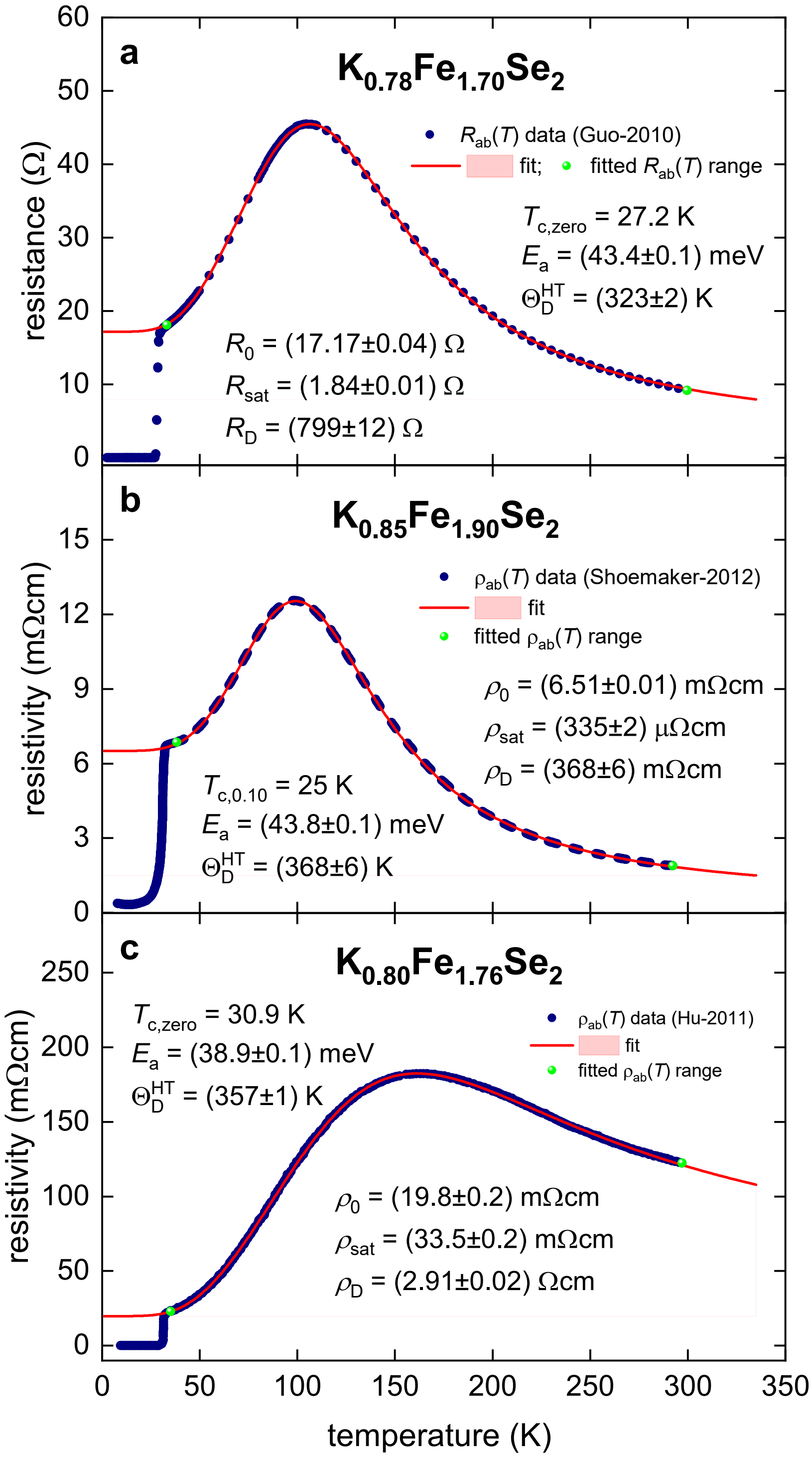

**Figure 4.** Fits with the nearest-neighbor hopping parallel resistance model (Equation 14) of the $\rho_{ab}(T)$ data measured in 122-type chalcogenide single crystals. (**a**) $K_{0.78}Fe_{1.70}Se_2$ (data reported by Guo et al. [5]). Fit quality *R*-square (COD) = 0.9999. (**b**) $K_{0.85}Fe_{1.90}Se_2$ (data reported by Shoemaker et al. [138]). Fit quality *R* = 0.9999. (**c**) $K_{0.80}Fe_{1.76}Se_2$ (data reported by Hu et al. [161]). Fit quality *R* = 0.9998. Deduced parameters are shown in each panel. The 95% confidence bands are shown by pink shadow areas.

### 3.4. Single crystals of the 111-type

Wang et al. [8] discovered superconductivity in the LiFeAs compound (111-type) with $T_c$ = 18 K. Two weeks later, Tapp et al. [162] confirmed the discovery, and several months later, Parker et al. [163] reported that another 111-type NaFeAs exhibits the same $T_c$ = 18 K. There is a low-$T_c$ member of the 111-type LiFeP [27], for which perfect single crystals were grown and studied by Kasahara et al. [164].

Morozov et al. [57] grew large LiFeAs single crystals with dimensions up to 12x12x0.3 mm$^3$ using the self-flux technique and reported $\rho_{ab}(T)$ data for one sample with $T_{c,zero} = 15.5$ K shown in Figure 5,a along with the fit using Equation 14. The deduced Debye temperature for this sample is $\Theta_D^{HT} = (298 \pm 1)$ K.

The same research group reported other $\rho_{ab}(T)$ dataset [165] for different LiFeAs single crystal with $T_{c,zero} = 17$ K, which is shown in Figure 5,b together with the fit using Equation 14. The deduced Debye temperature for that sample reported by Heyer et al. [165] is in the same range of values $\Theta_D^{HT} = (265 \pm 1)$ K.

The analysis of independently reported $\rho_{ab}(T)$ data for sample FP1 grew by Rullier-Albenque et al. [104,166] is shown in Figure 5,c. Samples grew up by this group have very high residual resistivity ratios RRR $= \frac{\rho_{ab}(T=300\text{ K})}{\rho_0} \cong 250$ [104,166]. The deduced Debye temperature for FP1 is in the same range of values $\Theta_D^{HT} = (301 \pm 1)$ K.

The Aslamazov-Larkin analysis [147] of $\rho_{ab}(T)$ near $T_c$ performed by Rullier-Albenque et al. [104] for LiFeAs samples showed that this phase exhibits a high level of 2D fluctuations, and the excess conductivity appeared well above the transition temperature. Based on this, data fittings using Equation 14 for LiFeAs (in Figure 4) were performed for $\rho_{ab}(T \geq 2 \times T_{c,zero})$.

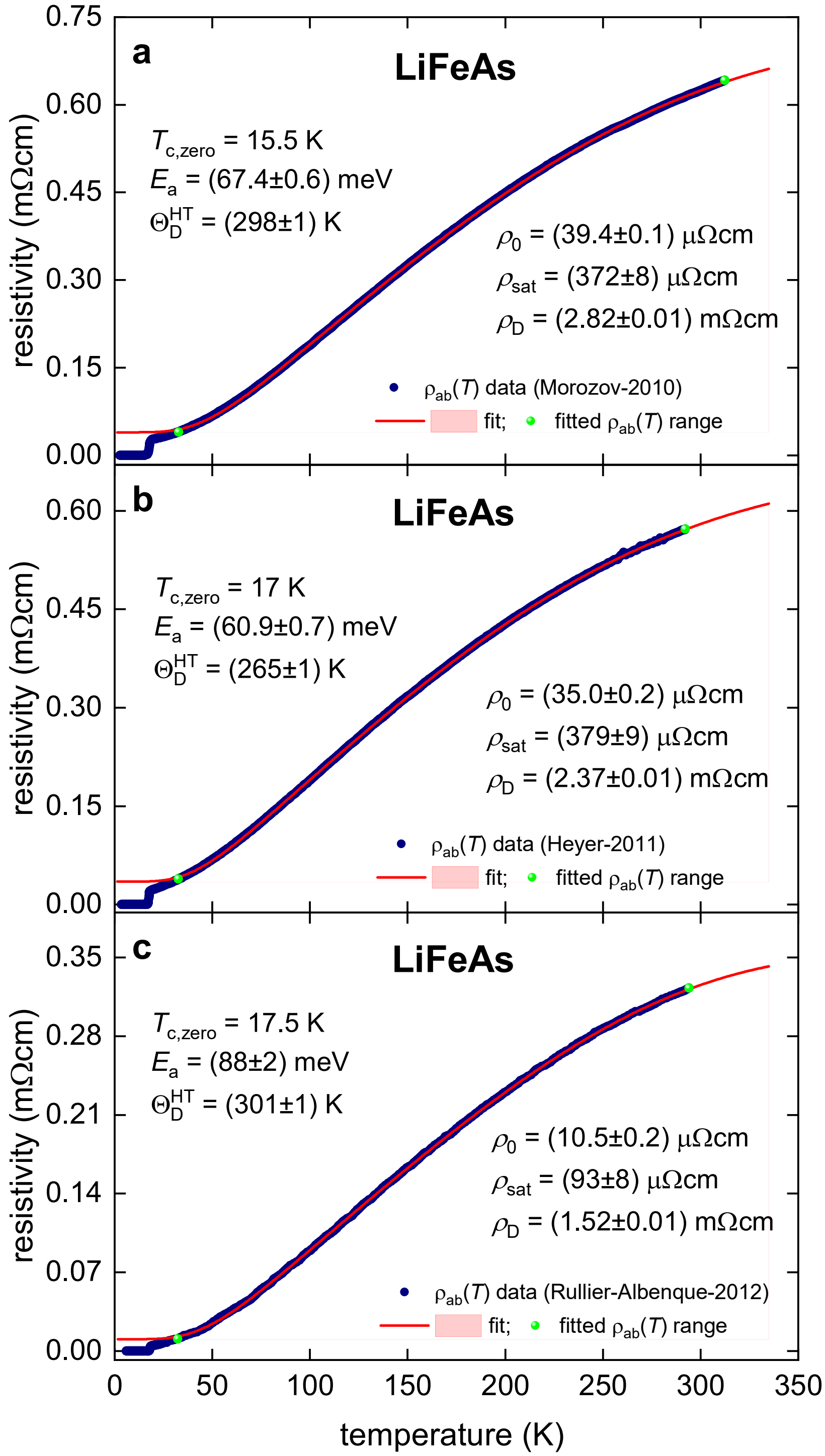

**Figure 5.** Fits with the nearest-neighbor hopping parallel resistance model (Equation 14) of the $\rho_{ab}(T)$ data measured in 111-type single crystals LiFeAs. (**a**) data reported by Morozov et al. [57]). Fit quality *R*-square (COD) = 0.9999. (**b**) Data reported by Heyer et al. [165]. Fit quality *R* = 0.9999. (**c**) Data reported by Rullier-Albenque et al. [104,166] for sample FP1. Fit quality *R* = 0.9999. Deduced parameters are shown in each panel. The 95% confidence bands are shown by pink shadow areas.

### 3.5. Single crystals of the 1144-type

Iyo et al. [4] discovered superconductivity in the 1144-type of IBS. In particular, Iyo et al. [4] reported that ceramics of $CaAFe_4As_4$ (A = K, Rb, Cs) and $SrAFe_4As_4$ (A = Rb, Cs) exhibit $31\ \mathrm{K} \leq T_{\mathrm{c,zero}} \leq 37$ K. There are interesting compounds of this 1144-type with europium $AEuFe_4As_4$ (A = Rb, Sc), in which superconductivity coexists with ferromagnetism at low temperatures [16,167].

Bao et al. [168] grew $RbEuFe_4As_4$ single crystals and reported $\rho_{\mathrm{ab}}(T)$ data for a sample with $T_{\mathrm{c,zero}} = 36.8$ K. The dataset, together with the fit using Equation 14, is shown in Figure 6,a. The deduced Debye temperature for this sample is $\Theta_{\mathrm{D}}^{\mathrm{HT}} = (243 \pm 2)$ K.

Bristow et al. [169] grew $CaKFe_4As_4$ single crystals and reported $\rho_{\mathrm{ab}}(T)$ data for a sample S2 with $T_{\mathrm{c,zero}} = 34.9$ K. The dataset and the fit are shown in Figure 6,b. Primary deduced parameters for this sample are $\Theta_{\mathrm{D}}^{\mathrm{HT}} = (272 \pm 1)$ K and $E_{\mathrm{a}} = (25.3 \pm 0.2)$ meV. These parameters are practically the same as the parameters deduced by the analysis of the $\rho_{\mathrm{ab}}(T)$ data reported for the $CaKFe_4As_4$ single crystal by an independent research group, Pyon et al. [170] with $T_{\mathrm{c,zero}} = 35.3$ K. The results of the analysis for data reported by Pyon et al. [170] is shown in Figure 6,c. It should also be noted that the absolute values of the $\rho_{\mathrm{ab}}(T)$ in studies [169,170] differs by a factor of two.

This result can be served as evidence for the nearest-neighbours hopping parallel resistivity model reliability.

There are two main found differences between deduced parameters for $RbEuFe_4As_4$ and $CaKFe_4As_4$. One is that the $E_{\mathrm{a}}$ in the RbEu-based crystal is nearly twice larger that the value in the CaK-based crystals, while the Debye temperature in the RbEu-based crystal is ~ 20% lower than this in the CaK-based counterparts.

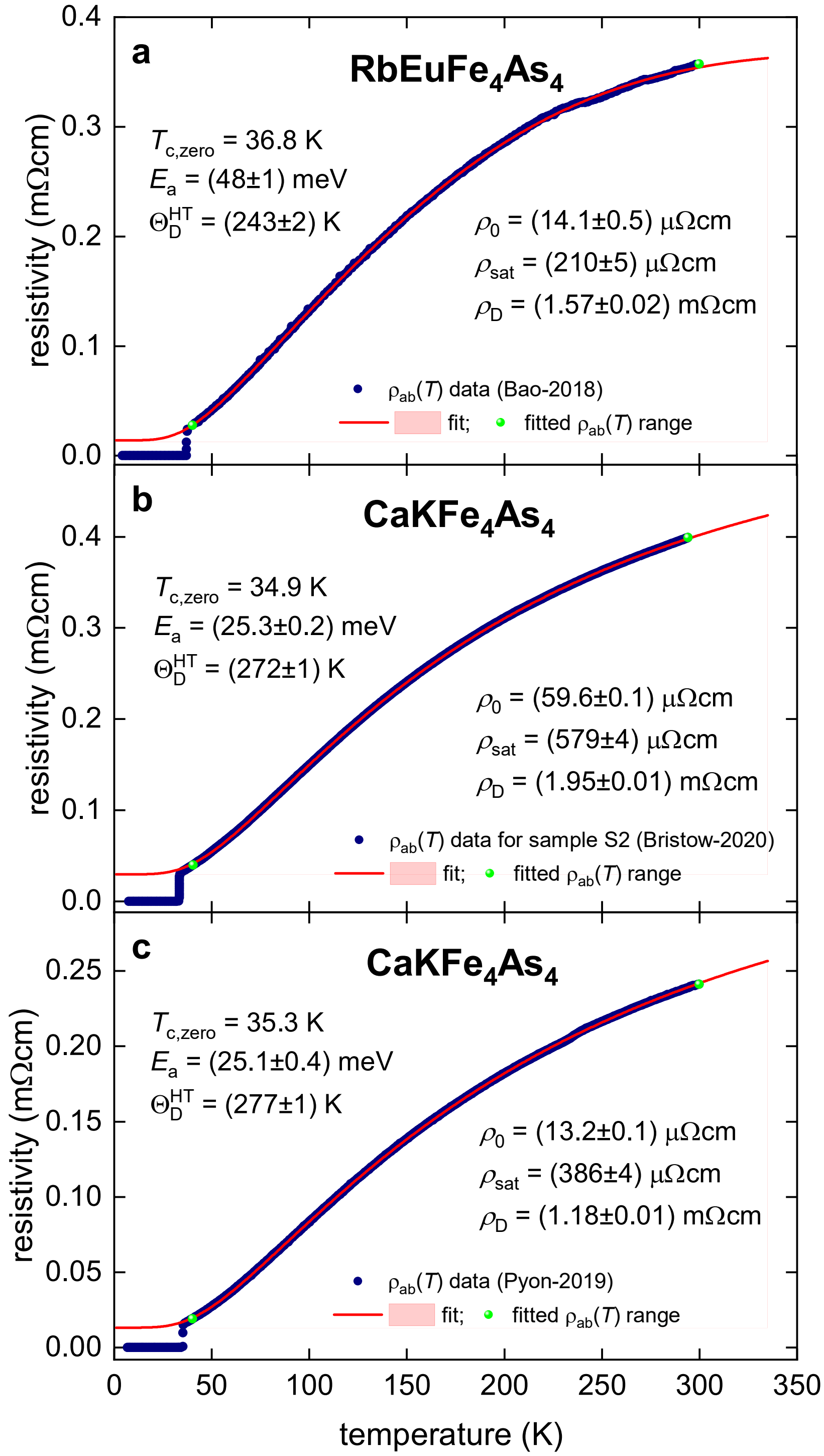

**Figure 6.** Fits with the nearest-neighbor hopping parallel resistance model (Equation 14) of the $\rho_{ab}(T)$ data measured in 1144-type single crystals. (**a**) Data for the $RbEuFe_4As_4$ single crystal reported by Bao et al. [168]). Fit quality *R*-square (COD) = 0.9999. (**b**) Data for the $CaKFe_4As_4$ single crystal (S2) reported by Bristow et al. [169]). Fit quality *R* = 0.9999. (**c**) Data for the $CaKFe_4As_4$ single crystal reported by Pyon et al. [170]. Fit quality *R* = 0.9999. Deduced parameters are shown in each panel. The 95% confidence bands are shown by pink shadow areas.

### 3.6. Single crystal of the 42214-type

Katrych et al. [17,171] discovered superconductivity in the 42214-type of IBS compounds $L_4Fe_2As_2Te_{1-x}O_{1-y}F_y$ ($L$=Pr, Sm, Gd) that exhibit the highest transition temperature $T_{c,magnetization} \cong 45$ K measured by DC magnetization.

Pisoni et al. [135] grew $Sm_4Fe_2As_2Te_{0.72}O_{2.8}F_{1.2}$ single crystals at high temperature and high pressure employing a cubic-anvil system. The reported $\rho_{ab}(T)$ data for the sample with $T_{c,zero} = 31.3$ K together with the fit using Equation 14, is shown in Figure 7. The deduced Debye temperature for this sample is $\Theta_D^{HT} = (330 \pm 2)$ K.

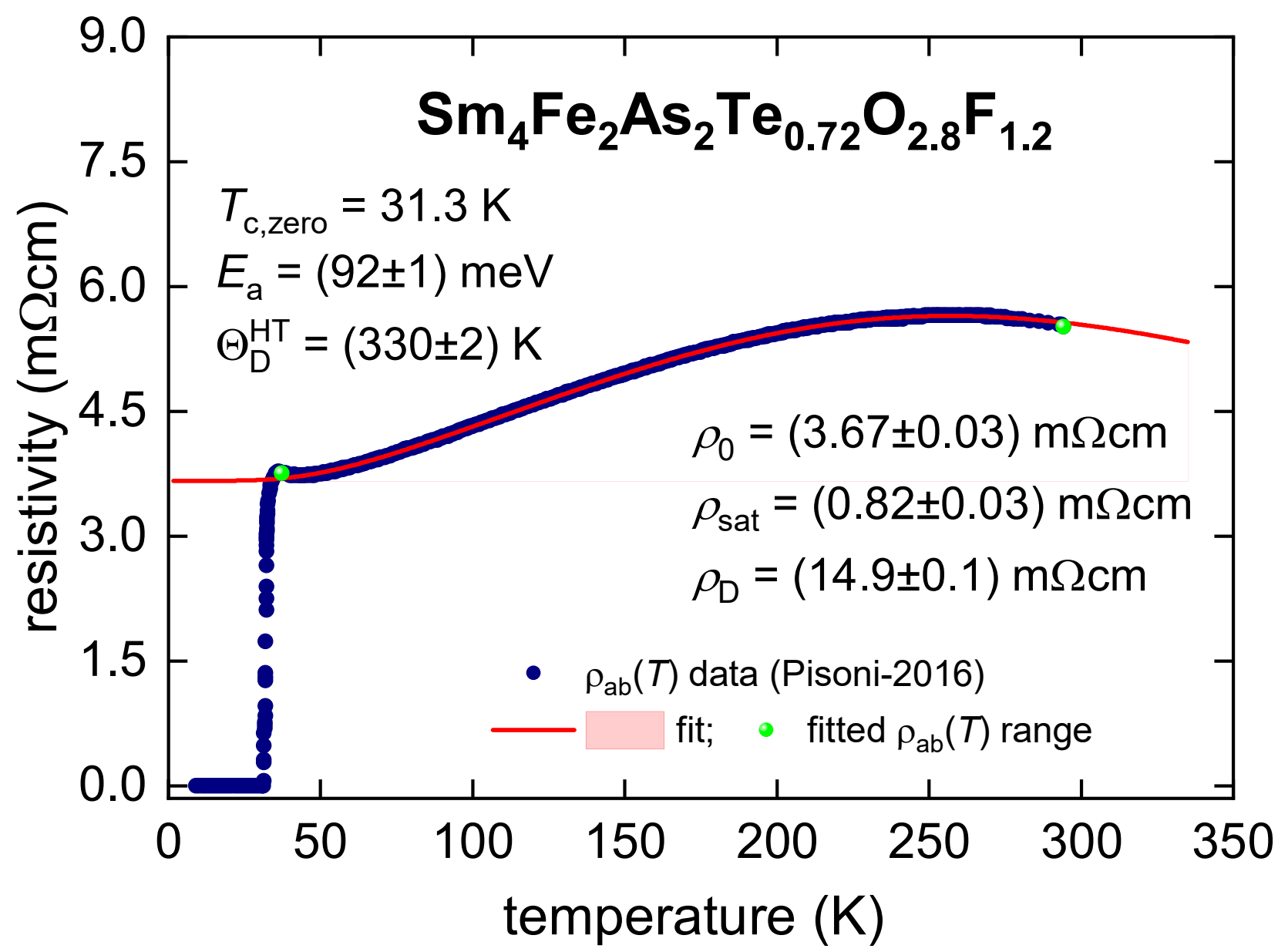

**Figure 7.** Fit with the nearest-neighbor hopping parallel resistance model (Equation 14) of the $\rho_{ab}(T)$ data measured in the 42214-type single crystal. Data for the $Sm_4Fe_2As_2Te_{0.72}O_{2.8}F_{1.2}$ single crystal reported by Pisoni et al. [135]). Fit quality $R$-square (COD) = 0.9998. Deduced parameters are shown in the figure. The 95% confidence bands are shown by pink shadow areas.

### 3.7. Single crystals of the 11-type

Hsu et al. [9] discovered superconductivity in FeSe exhibiting $T_{c,zero} = 7$ K at room pressure, and Medvedev et al. [172] showed that the transition temperature increases rapidly under pressure and reaches $T_c = 36.7$ K at $P = 8.9$ GPa. Atomically thin epitaxial FeSe films have an even higher transition temperature [173–176].

Since bulk pure FeSe undergoes a structural nematic transition at $T_s \sim 90$ K, the analysis of $\rho_{ab}(T)$ for this type of IBS is beyond the scope of the present study (and the analysis will be provided in a separate report), I have presented here the analysis for $FeTe_{1-x}Se_x$ single crystals, where this structural transition is suppressed, and the $FeTe_{1-x}Se_x$ compounds are also 11-type of IBS.

Pallecchi et al. [177] grew iron-excess $Fe_{1+y}Te_{1-x}Se_x$ single crystals by the Bridgeman method. The $\rho_{ab}(T)$ data reported for the sample with $T_{c,zero} = 12$ K and the composition of $Fe_{1.013}Te_{1-x}Se_x$ (x=0.3) data shown in Figure 8,a together with the fit using Equation 14. The deduced Debye temperature for this sample is $\Theta_D^{HT} = (249 \pm 12)$ K.

Liu et al. [178] grew $Fe_{1+y}Te_{1-x}Se_x$ single crystals, and the $\rho_{ab}(T)$ data reported for the sample SC1 with $T_{c,zero} = 13.2$ K and the measured average composition $Fe_{1.03}Te_{1-x}Se_x$ (x=0.37) are shown in Figure 8,b together with the fit using Equation 14. The deduced Debye temperature for this sample is $\Theta_D^{HT} = (176 \pm 9)$ K.

Sun et al. [179] grew $Fe_{1+y}Te_{1-x}Se_x$ single crystals by the self-flux method [180], and the $\rho_{ab}(T)$ data reported for the sample with $T_{c,zero} = 14.2$ K and the composition of $FeTe_{1-x}Se_x$ (x=0.43) are shown in Figure 8,c together with the fit using Equation 14. The deduced Debye temperature for this sample is $\Theta_D^{HT} = (172 \pm 6)$ K.

It should be noted that despite the deduced Debye temperature varying in some range, $170 \text{ K} \leq \Theta_D^{HT} \leq 250$ K, all analyzed $Fe_{1+y}Te_{1-x}Se_x$ single crystals exhibit practically the same Arrhenius activation energy $E_a \cong 40$ meV.

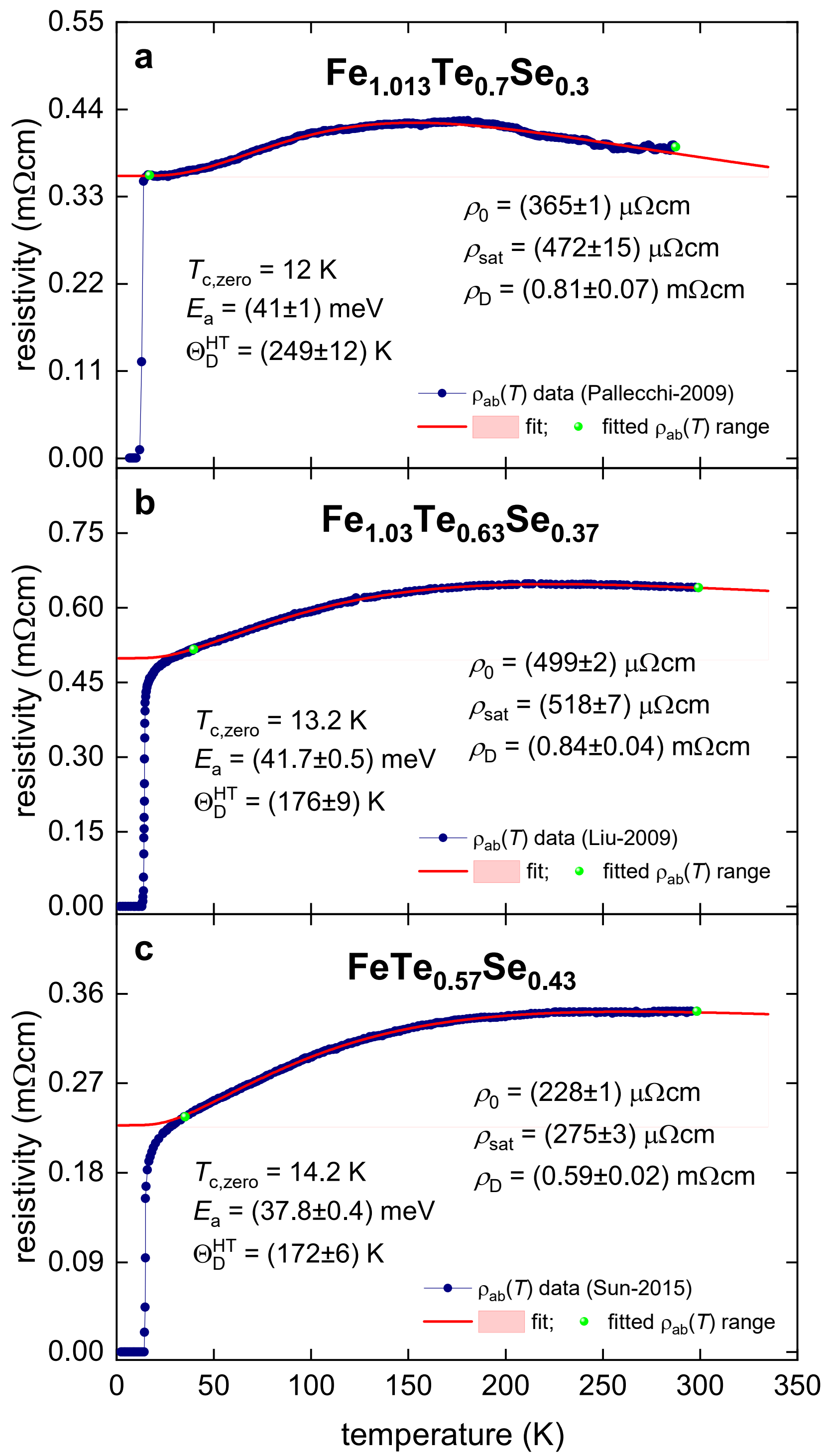

**Figure 8.** Fits with the nearest-neighbor hopping parallel resistance model (Equation 14) of the $\rho_{ab}(T)$ data measured in 11-type single crystals. (**a**) Data for the $Fe_{1.013}Te_{1-x}Se_x$ (x=0.3) single crystal reported by Pallecchi et al. [177]). Fit quality *R*-square (COD) = 0.9854. (**b**) Data for the $Fe_{1.03}Te_{1-x}Se_x$ (x=0.37) single crystal reported by Liu et al. [169]). Fit quality *R* = 0.9999. (**c**) Data for the $CaKFe_4As_4$ single crystal reported by Pyon et al. [170]. Fit quality *R* = 0.9999. Deduced parameters are shown in each panel. The 95% confidence bands are shown by pink shadow areas.

## 4. Discussion

Debye temperature $\Theta_D$ values have been reported for virtually all IBS superconductors. However, in many reports the $\Theta_D$ values were obtained by approximating the low-temperature heat capacity $C_p(T)$ data using Equation 6 and subsequent calculations based on the derived parameter $\beta_3$ using Equation 7.

However, as it was pointed out by McQueen et al. [130] and Tsurkan et al. [181] , this approach "…*is generally only good up to* $\frac{\Theta_D}{50}$ … " [130]. Considering that the ballpark value for $\Theta_D$ in IBS is about $\Theta_D = 250$ K, the reported values for $\Theta_D$ in IBS are to obey the inequality:

$$\frac{\Theta_D}{50} \cong 5\ K \ll T_c, \tag{17}$$

and thus, these values need to be designated as $\Theta_D(T \cong 5\ K) \equiv \Theta_D^{LT}$ (this clarification was also pointed out by Nowak et al. [127] in the case of high-entropy alloys).

In this study, the Debye temperatures $\Theta_D^{HT}$ were obtained from experimental data measured in the range:

$$1.5 \times T_{c,zero} \lesssim T \leq 300\ \text{K}, \tag{18}$$

and thus, the deduced $\Theta_D^{HT}$ values should be close to the values $\Theta_D^{HT}$ reported by the Mössbauer spectroscopy technique [182–184].

Since the temperature range indicated in Equation 18 is closer to and precedes the observed $T_c$, the obtained $\Theta_D^{HT}$ values are more appropriate for testing the possibility of classical electron-phonon mechanisms for superconductivity in IBS (which may be assisted, enhanced, or operating in conjunction with other mechanisms). A possibility for the electron-phonon mediated superconductivity in IBS is under ongoing discussion since the discovery of these superconductors.

I calculated the electron-phonon coupling constant $\lambda_{\mathrm{e-p}}$ for all the superconductors studied (Table I) by using the following equations [119,185,186] (where it was assumed that $\mu^* = 0.125$):

$$T_{\mathrm{c,zero}} = \frac{\Theta_{\mathrm{D}}^{\mathrm{HT}}}{1.45} e^{\left[-\frac{1.04\times(1+\lambda_{e-ph})}{\lambda_{e-ph}-\mu^*\times(1+0.62\times\lambda_{e-ph})}\right]} \times f_1 \times f_2^* \qquad (19)$$

$$f_1 = \left(1 + \left(\frac{\lambda_{e-ph}}{2.46(1+3.8\mu^*)}\right)^{3/2}\right)^{1/3} \qquad (20)$$

$$f_2^* = 1 + (0.0241 - 0.0735 \times \mu^*) \times \lambda_{e-ph}^2 \qquad (21)$$

The obtained values of $\lambda_{\mathrm{e-ph}}$ (Table I) show that the electron-phonon mechanism as the main mechanism of superconductivity in IBS can be considered as a possibility.

## 5. Conclusions

In this study, it is shown that the in-plane resistivity data $\rho_{\mathrm{ab}}(T)$ in seven types of iron-based superconductors, namely, 11, 111, 122-pnictides, 122-chalcogenides, 1111, 1144 and 42214, are well described by the nearest-neighbor hopping parallel resistance model [141], the basic idea of which was proposed by Shoemaker et al. [138], and the complete physical equation for the temperature-dependent metallic conductivity channel was proposed in [141].

**Table I**. Calculated the electron-phonon coupling constant $\lambda_{\mathrm{e-ph}}$ for all studied superconductors. Equations 19-21 were used for calculations with the assumed $\mu^* = 0.125$. $\Theta_{\mathrm{D}}$ is the Debye temperature, $\Theta_{\mathrm{E}}$ is the Einstein temperature.

| IBS type | Chemical formula | Observed $T_{c,\mathrm{zero}}$ (K) | Reported $\Theta_{\mathrm{D}}^{\mathrm{LT}}$ (K) | Reported $\Theta_{\mathrm{D}}^{\mathrm{HT}}$ (K) | Derived $\Theta_{\mathrm{D}}^{\mathrm{HT}}$ (K) | $\lambda_{\mathrm{e-ph}}$ |
|---|---|---|---|---|---|---|
| 11 | $Fe_{1.013}Te_{1-x}Se_x$ (x=0.3) [177] | 12 [177] | 174 [187]<br>200 [130]<br>235 [181]<br>240 [182] | 285 [182]<br>$\Theta_{\mathrm{D,1}}^{\mathrm{HT}} = 127$ [181]<br>$\Theta_{\mathrm{D,2}}^{\mathrm{HT}} = 235$ [181]<br>$\Theta_{\mathrm{E}}^{\mathrm{HT}} = 315$ [181] | $249 \pm 12$ | 0.94 |
| | $Fe_{1.03}Te_{1-x}Se_x$ (x=0.37) [178] | 13.2 [178] | | | $176 \pm 9$ | 1.23 |
| | $FeTe_{1-x}Se_x$ (x=0.43) [179] | 14.2 [179] | | | $172 \pm 6$ | 1.32 |
| | | | | | | |
| 111 | LiFeAs [57] | 15.5 [57] | 310 [188]<br>300 [189]<br>290 for NaFeAs [190] | | $298 \pm 1$ | 0.98 |
| | LiFeAs [165] | 17 [165] | | | $265 \pm 1$ | 1.11 |
| | LiFeAs [104,166] | 17.5 [104] | | | $301 \pm 1$ | 1.04 |
| | | | | | | |
| 122 pnic-tides | $(Ba_{1-x}K_x)Fe_2As_2$ (x=0.4) [159] | 36 [159] | 274 [191]<br>260 [192]<br>246 (x = 0.5) [193]<br>250 for $La_{0.4}Na_{0.6}Fe_2As_2$ [131]<br>292 for $CaFe_2As_2$ [194] | ~ 292 for $CaFe_2As_2$ [194] | $306 \pm 1$ | 1.81 |
| | $(Ba_{1-x}K_x)Fe_2As_2$ (x=0.4) [160] | 37 [160] | | | $305 \pm 7$ | 1.86 |
| | | | | | | |
| 122 chalco-genides | $K_{0.78}Fe_{1.70}Se_2$ [5] | 27.2 [5] | 212 [195]<br>197 [196] | | $323 \pm 2$ | 1.34 |
| | $K_{0.85}Fe_{1.90}Se_2$ [138] | 25 [138] | | | $368 \pm 6$ | 1.15 |
| | $K_{0.80}Fe_{1.76}Se_2$ [161] | 30.9 [161] | | | $357 \pm 1$ | 1.37 |
| | $(K_{1-x}Na_x)_yFe_{1.75}Se_2$ (x=0.32; y = 0.95) [140] | 31.7 [140] | | | $536 \pm 1$ | 1.05 |
| | | | | | | |
| 1111 | $Ca(Fe_{1-x}Co_x)AsF$ (x = 0.118) [155] | 19 [155] | 339 for SrFeAsF [197]<br>285 for $PrFeAsO_{1-x}F_x$ [198]<br>316 for $LaFeAsO_{0.9}F_{0.1}$ [199]<br>323 for LaFeAsO [132] | 342 for SrFeAsF [200]<br>347 for SrFeAsF [200]<br>294 for $PrFeAsO_{1-x}F_x$ [198]<br>355 [201] | $599 \pm 6$ | 0.78 |
| | $NdFeAsO_{1-x}F_x$ (x=0.1) [53] | 40 [53] | | | $449 \pm 2$ | 1.40 |
| | $La_{0.87}Sm_{0.13}FeAs_{0.91}P_{0.09}O$ [156] | 13.4 [156] | | | $243 \pm 3$ | 1.01 |
| | | | | | | |
| 1144 | $RbEuFe_4As_4$ [168] | 36.8 [168] | | 391 [184] | $243 \pm 2$ | 2.38 |
| | $CaKFe_4As_4$ [169] | 34.9 [169] | | | $272 \pm 1$ | 1.98 |
| | $CaKFe_4As_4$ [170] | 35.5 [170] | | | $277 \pm 1$ | 1.98 |
| | | | | | | |
| 42214 | $Sm_4Fe_2As_2Te_{0.72}O_{2.8}F_{1.2}$ | 31.3 | | | $330 \pm 2$ | 1.48 |

**Data availability statement**

The data that support the findings of this study are available from the corresponding author upon reasonable request.

**Declaration of interests**

The author declares that he has no known competing financial interests or personal relationships that could have appeared to influence the work reported in this paper.

**Acknowledgement**

The work was carried out within the framework of the state assignment of the Ministry of Science and Higher Education of the Russian Federation for the IMP UB RAS. The author gratefully acknowledges the research funding from the Ministry of Science and Higher Education of the Russian Federation under the Ural Federal University Program of Development within the Priority-2030 Program.

**References**

[1] Y. Kamihara, H. Hiramatsu, M. Hirano, R. Kawamura, H. Yanagi, T. Kamiya, and H. Hosono, Iron-Based Layered Superconductor: LaOFeP, J. Am. Chem. Soc. **128**, 10012 (2006).

[2] H. Takahashi, K. Igawa, K. Arii, Y. Kamihara, M. Hirano, and H. Hosono, Superconductivity at 43 K in an iron-based layered compound LaO1-xFxFeAs, Nature **453**, 376 (2008).

[3] Y. Kamihara, T. Watanabe, M. Hirano, and H. Hosono, Iron-Based Layered Superconductor La[$O_{1-x}F_x$]FeAs ($x$ = 0.05−0.12) with $T_c$ = 26 K, J. Am. Chem. Soc. **130**, 3296 (2008).

[4] A. Iyo, K. Kawashima, T. Kinjo, T. Nishio, S. Ishida, H. Fujihisa, Y. Gotoh, K. Kihou, H. Eisaki, and Y. Yoshida, New-Structure-Type Fe-Based Superconductors: $CaAFe_4As_4$ ($A$ = K, Rb, Cs) and $SrAFe_4As_4$ ($A$ = Rb, Cs), J. Am. Chem. Soc. **138**, 3410 (2016).

[5] J. Guo, S. Jin, G. Wang, S. Wang, K. Zhu, T. Zhou, M. He, and X. Chen, Superconductivity in the iron selenide KxFe2Se2 (0 ≤ x ≤ 1.0), Phys. Rev. B **82**, 180520 (2010).

[6] M. Rotter, M. Tegel, and D. Johrendt, Superconductivity at 38 K in the Iron Arsenide (Ba(1−x)K(x))Fe2As2, Phys. Rev. Lett. **101**, 107006 (2008).

[7] M. Rotter, M. Pangerl, M. Tegel, and D. Johrendt, Superconductivity and Crystal Structures of (Ba(1−x)K(x))Fe 2As 2 ( x =0–1), Angew. Chem. Int. Ed. **47**, 7949 (2008).

[8] X. C. Wang, Q. Q. Liu, Y. X. Lv, W. B. Gao, L. X. Yang, R. C. Yu, F. Y. Li, and C. Q. Jin, The superconductivity at 18 K in LiFeAs system, Solid State Commun. **148**, 538 (2008).

[9] F.-C. Hsu et al., Superconductivity in the PbO-type structure α-FeSe, Proceedings of the National Academy of Sciences **105**, 14262 (2008).

[10] M. H. Fang, H. M. Pham, B. Qian, T. J. Liu, E. K. Vehstedt, Y. Liu, L. Spinu, and Z. Q. Mao, Superconductivity close to magnetic instability in FeSe1−xTe0.82, Phys. Rev. B **78**, 224503 (2008).

[11] S. Kakiya, K. Kudo, Y. Nishikubo, K. Oku, E. Nishibori, H. Sawa, T. Yamamoto, T. Nozaka, and M. Nohara, Superconductivity at 38 K in Iron-Based Compound with Platinum–Arsenide Layers $Ca_{10}(Pt_4As_8)(Fe_{2-x}Pt_xAs_2)_5$, J. Physical Soc. Japan **80**, 093704 (2011).

[12] N. Ni, J. M. Allred, B. C. Chan, and R. J. Cava, High $T_c$ electron doped $Ca_{10}(Pt_3As_8)(Fe_2As_2)_5$ and $Ca_{10}(Pt_4As_8)(Fe_2As_2)_5$ superconductors with skutterudite intermediary layers, Proceedings of the National Academy of Sciences **108**, (2011).

[13] Y.-L. Sun, H. Jiang, H.-F. Zhai, J.-K. Bao, W.-H. Jiao, Q. Tao, C.-Y. Shen, Y.-W. Zeng, Z.-A. Xu, and G.-H. Cao, $Ba_2Ti_2Fe_2As_4O$: A New Superconductor Containing $Fe_2As_2$ Layers and $Ti_2O$ Sheets, J. Am. Chem. Soc. **134**, 12893 (2012).

[14] S. Fujitsu, S. Matsuishi, and H. Hosono, Iron based superconductors processing and properties, International Materials Reviews **57**, 311 (2012).

[15] N. D. Zhigadlo, S. Weyeneth, S. Katrych, P. J. W. Moll, K. Rogacki, S. Bosma, R. Puzniak, J. Karpinski, and B. Batlogg, High-pressure flux growth, structural, and superconducting properties of LnFeAsO ( Ln= Pr, Nd, Sm) single crystals, Phys. Rev. B **86**, 214509 (2012).

[16] Y. Liu et al., Superconductivity and ferromagnetism in hole-doped RbEuFe4As4, Phys. Rev. B **93**, 214503 (2016).

[17] S. Katrych, A. Pisoni, S. Bosma, S. Weyeneth, N. D. Zhigadlo, R. Gaal, J. Karpinski, and L. Forró, L4Fe2As2Te1−xO4−yFy (L = Pr, Sm, Gd): A layered oxypnictide superconductor with Tc up to 45 K, Phys. Rev. B **89**, 024518 (2014).

[18] J. G. Bednorz and K. A. Müller, Possible high T c superconductivity in the Ba-La-Cu-O system, Zeitschrift Für Physik B Condensed Matter **64**, 189 (1986).

[19] J. Nagamatsu, N. Nakagawa, T. Muranaka, Y. Zenitani, and J. Akimitsu, Superconductivity at 39 K in magnesium diboride, Nature **410**, 63 (2001).

[20] A. P. Drozdov, M. I. Eremets, I. A. Troyan, V. Ksenofontov, and S. I. Shylin, Conventional superconductivity at 203 kelvin at high pressures in the sulfur hydride system, Nature **525**, 73 (2015).

[21] D. Li, K. Lee, B. Y. Wang, M. Osada, S. Crossley, H. R. Lee, Y. Cui, Y. Hikita, and H. Y. Hwang, Superconductivity in an infinite-layer nickelate, Nature **572**, 624 (2019).

[22] H. Sun et al., Signatures of superconductivity near 80 K in a nickelate under high pressure, Nature **621**, 493 (2023).

[23] R. J. Cava, B. Batlogg, J. J. Krajewski, R. Farrow, L. W. Rupp, A. E. White, K. Short, W. F. Peck, and T. Kometani, Superconductivity near 30 K without copper: the Ba0.6K0.4BiO3 perovskite, Nature **332**, 814 (1988).

[24] H. Hosono and K. Kuroki, Iron-based superconductors: Current status of materials and pairing mechanism, Physica C: Superconductivity and Its Applications **514**, 399 (2015).

[25] Y. Mizuguchi, F. Tomioka, S. Tsuda, T. Yamaguchi, and Y. Takano, Substitution Effects on FeSe Superconductor, J. Physical Soc. Japan **78**, 074712 (2009).

[26] Z. Zajicek, S. J. Singh, H. Jones, P. Reiss, M. Bristow, A. Martin, A. Gower, A. McCollam, and A. I. Coldea, Drastic effect of impurity scattering on the electronic and superconducting properties of Cu-doped FeSe, Phys. Rev. B **105**, 115130 (2022).

[27] Z. Deng, X. C. Wang, Q. Q. Liu, S. J. Zhang, Y. X. Lv, J. L. Zhu, R. C. Yu, and C. Q. Jin, A new "111" type iron pnictide superconductor LiFeP, EPL (Europhysics Letters) **87**, 37004 (2009).

[28] K. Iida, J. Hänisch, and C. Tarantini, Fe-based superconducting thin films on metallic substrates: Growth, characteristics, and relevant properties, Appl. Phys. Rev. **5**, 031304 (2018).

[29] M. Sakoda, K. Iida, and M. Naito, Recent progress in thin-film growth of Fe-based superconductors: superior superconductivity achieved by thin films, Supercond. Sci. Technol. **31**, 093001 (2018).

[30] P. Cayado, J. Hänisch, K. Iida, and C. Senatore, Focus on recent advances in superconducting films, Supercond. Sci. Technol. **36**, 090401 (2023).

[31] K. Iida, J. Hänisch, S. Hata, and A. Yamamoto, Recent progress on epitaxial growth of Fe-based superconducting thin films, Supercond. Sci. Technol. **36**, 063001 (2023).

[32] J. Hänisch, K. Iida, R. Hühne, and C. Tarantini, Fe-based superconducting thin films—preparation and tuning of superconducting properties, Supercond. Sci. Technol. **32**, 093001 (2019).

[33] Y. Tsuji, T. Hatano, K. Kondo, K. Iida, Y. Mawatari, and H. Ikuta, Microfabrication of NdFeAs(O,F) thin films and evaluation of the transport properties, Supercond. Sci. Technol. **33**, 074001 (2020).

[34] M. Chihara, N. Sumiya, K. Arai, A. Ichinose, I. Tsukada, T. Hatano, K. Iida, and H. Ikuta, Direct growth of superconducting NdFeAs(O,F) thin films by MBE, Physica C: Superconductivity and Its Applications **518**, 69 (2015).

[35] J. Hänisch et al., High field superconducting properties of Ba(Fe1−xCox)2As2 thin films, Sci. Rep. **5**, 17363 (2015).

[36] E. F. Talantsev, K. Iida, T. Ohmura, T. Matsumoto, W. P. Crump, N. M. Strickland, S. C. Wimbush, and H. Ikuta, p-wave superconductivity in iron-based superconductors, Sci. Rep. **9**, 14245 (2019).

[37] T. E. Kuzmicheva, S. A. Kuzmichev, I. V. Morozov, S. Wurmehl, and B. Büchner, Experimental Evidence of Three-Gap Superconductivity in LiFeAs, JETP Lett. **111**, 350 (2020).

[38] S. A. Kuzmichev, T. E. Kuzmicheva, A. I. Boltalin, and I. V. Morozov, Multiple andreev reflections spectroscopy of superconducting LiFeAs single crystals: Anisotropy and temperature behavior of the order parameters, JETP Lett. **98**, 722 (2014).

[39] I. I. Mazin, D. J. Singh, M. D. Johannes, and M. H. Du, Unconventional Superconductivity with a Sign Reversal in the Order Parameter of LaFeAsO1−xFx, Phys. Rev. Lett. **101**, 057003 (2008).

[40] L. Zhao et al., Common electronic origin of superconductivity in (Li,Fe)OHFeSe bulk superconductor and single-layer FeSe/SrTiO3 films, Nat. Commun. **7**, 10608 (2016).

[41] O. Chmaissem, R. Stadel, K. M. Taddei, D. Bugaris, D. D. Khalyavin, P. Manuel, D. Y. Chung, M. G. Kanatzidis, R. Osborn, and S. Rosenkranz, Disorder-induced universality and scaling in hole-doped iron-based superconductors, Commun. Mater. **6**, 146 (2025).

[42] P. J. Hirschfeld, M. M. Korshunov, and I. I. Mazin, Gap symmetry and structure of Fe-based superconductors, Reports on Progress in Physics **74**, 124508 (2011).

[43] D. C. Johnston, The puzzle of high temperature superconductivity in layered iron pnictides and chalcogenides, Adv. Phys. **59**, 803 (2010).

[44] Z. Ren and Z. Zhao, Research and Prospects of Iron-Based Superconductors, Advanced Materials **21**, 4584 (2009).

[45] Y. A. Izyumov and E. Z. Kurmaev, FeAs systems: a new class of high-temperature superconductors, Physics-Uspekhi **51**, 1261 (2008).

[46] Y. Izyumov and E. Kurmaev, *High-Tc Superconductors Based on FeAs Compounds*, Vol. 143 (Springer Berlin Heidelberg, Berlin, Heidelberg, 2010).

[47] T. K. Kim et al., Electronic structure and coexistence of superconductivity with magnetism in RbEuFe4As4, Phys. Rev. B **103**, 174517 (2021).

[48] K. S. Pervakov, V. A. Vlasenko, E. P. Khlybov, A. Zaleski, V. M. Pudalov, and Y. F. Eltsev, Bulk magnetization and strong intrinsic pinning in Ni-doped BaFe2As2 single crystals, Supercond. Sci. Technol. **26**, 015008 (2013).

[49] T. E. Kuzmicheva et al., Superconducting gap symmetry in the superconductor BaFe1.9Ni0.1As2, Phys. Rev. B **97**, 235106 (2018).

[50] S. Kawachi, J. Yamaura, Y. Kuramoto, S. Iimura, T. Nomura, Y. Kohama, T. Sasaki, M. Tokunaga, Y. Murakami, and H. Hosono, Distinctive doping dependence of upper critical field in iron-based superconductor <math> <mrow> <msub> <mi>LaFeAsO</mi> <mrow> <mn>1</mn> <mo>−</mo> <mi>x</mi> </mrow> </msub> <msub> <mi mathvariant="normal">H</mi> <mi>x</mi> </msub> </mrow> </math>, Phys. Rev. B **108**, L100503 (2023).

[51] J. Paglione and R. L. Greene, High-temperature superconductivity in iron-based materials, Nat. Phys. **6**, 645 (2010).

[52] R. M. Fernandes, A. I. Coldea, H. Ding, I. R. Fisher, P. J. Hirschfeld, and G. Kotliar, Iron pnictides and chalcogenides: a new paradigm for superconductivity, Nature **601**, 35 (2022).

[53] A. Adamski, C. Krellner, and M. Abdel-Hafiez, Signature of multigap nodeless superconductivity in fluorine-doped NdFeAsO, Phys. Rev. B **96**, 100503 (2017).

[54] Q. Hou, L. Sun, Y. Sun, and Z. Shi, Review of Single Crystal Synthesis of 11 Iron-Based Superconductors, Materials **16**, 4895 (2023).

[55] L. Shlyk, K. K. Wolff, M. Bischoff, E. Rose, T. Schleid, and R. Niewa, Crystal structure and superconducting properties of hole-doped Ca $_{0.89}$ Na $_{0.11}$ FFeAs single crystals, Supercond. Sci. Technol. **27**, 044011 (2014).

[56] Mu Gang and Ma Yong-Hui, Single crystal growth and physical property study of 1111-type Fe-based superconducting system CaFeAsF, Acta Physica Sinica **67**, 177401 (2018).

[57] I. Morozov et al., Single Crystal Growth and Characterization of Superconducting LiFeAs, Cryst. Growth Des. **10**, 4428 (2010).

[58] A. V. Sadakov, V. A. Vlasenko, A. Yu. Levakhova, I. V. Zhuvagin, E. M. Fomina, V. A. Prudkoglyad, A. Y. Tsvetkov, A. S. Usoltsev, and N. D. Zhigadlo, Vortex matter and strong pinning in underdoped PrFeAs(O,F) with atomic-sized defects, Phys. Rev. Mater. **9**, 104801 (2025).

[59] M. Ishikado, S. Shamoto, H. Kito, A. Iyo, H. Eisaki, T. Ito, and Y. Tomioka, Growth of PrFeAsO1− single crystals and its characterization, Physica C: Superconductivity and Its Applications **470**, S322 (2010).

[60] S. L. Bud'ko, M. Sturza, D. Y. Chung, M. G. Kanatzidis, and P. C. Canfield, Heat capacity jump at Tc and pressure derivatives of superconducting transition temperature in the Ba1-xKxFe2As2 (0.2 ≤ x ≤ 1.0) series, Phys. Rev. B **87**, 100509 (2013).

[61] T. K. Kim et al., Novel magnetic stoichiometric superconductor compound EuRbFe<sub>4</sub>As<sub>4</sub>, Physics-Uspekhi **65**, 740 (2022).

[62] A. T. Danyarkhodzhaev, E. M. Ivanova, and A. S. Usoltsev, Multiband Superconductivity in DyFeAsO1–xFx, Bulletin of the Lebedev Physics Institute **53**, 67 (2026).

[63] Y. H. Kim, S. Park, C. Il Kwon, S. Y. Kim, K. Watanabe, T. Taniguchi, and J. S. Kim, Two-dimensional multiband superconductivity of the optimally and uniformly Li-intercalated FeSe nanoflakes, Current Applied Physics **46**, 27 (2023).

[64] D. Torsello, E. Piatti, G. A. Ummarino, X. Yi, X. Xing, Z. Shi, G. Ghigo, and D. Daghero, Nodal multigap superconductivity in the anisotropic iron-based compound RbCa2Fe4As4F2, NPJ Quantum Mater. **7**, 10 (2022).

[65] M. Putti et al., New Fe-based superconductors: properties relevant for applications, Supercond. Sci. Technol. **23**, 034003 (2010).

[66] P. Yang, H. Huang, M. Han, C. Liu, C. Yao, Y. Ma, and D. Wang, Iron-Based Superconductors for High-Field Applications: Realization of High Engineering Critical Current Density, Materials **17**, 5306 (2024).

[67] H. Ding et al., Development of the first Tesla class iron-based superconducting coil for high field application, Supercond. Sci. Technol. **36**, 11LT01 (2023).

[68] A. Yamamoto, A. Yamanaka, K. Iida, Y. Shimada, and S. Hata, Integrating machine learning with advanced processing and characterization for polycrystalline materials: a methodology review and application to iron-based superconductors, Sci. Technol. Adv. Mater. **26**, (2025).

[69] A. Mancini, A. Masi, A. Meledin, A. Angrisani Armenio, and G. Celentano, High $J_C$ behavior in novel TiN-buffered Fe(Se,Te) coated conductors driven by secondary phase nano-inclusions, Supercond. Sci. Technol. **39**, 015011 (2026).

[70] Y. Ma, Progress in wire fabrication of iron-based superconductors, Supercond. Sci. Technol. **25**, 113001 (2012).

[71] H. Hosono, A. Yamamoto, H. Hiramatsu, and Y. Ma, Recent advances in iron-based superconductors toward applications, Materials Today **21**, 278 (2018).

[72] X. Zhang and Y. Ma, Progress in the development of the 122-type IBS wires, Superconductivity **2**, 100010 (2022).

[73] M. Gurvitch and A. T. Fiory, Resistivity of La1.825Sr0.175CuO4 and YBa2Cu3O7 to 1100 K: Absence of saturation and its implications, Phys. Rev. Lett. **59**, 1337 (1987).

[74] A. P. Drozdov et al., Superconductivity at 250 K in lanthanum hydride under high pressures, Nature **569**, 528 (2019).

[75] P. Kong et al., Superconductivity up to 243 K in the yttrium-hydrogen system under high pressure, Nat. Commun. **12**, 5075 (2021).

[76] F. Du, A. P. Drozdov, V. S. Minkov, F. F. Balakirev, P. Kong, G. A. Smith, J. Yan, B. Shen, P. Gegenwart, and M. I. Eremets, Superconducting gap of H3S measured by tunnelling spectroscopy, Nature **641**, 619 (2025).

[77] V. S. Minkov et al., Long-term stability of superconducting metal superhydrides, Nat. Commun. **17**, 8171 (2026).

[78] T. Watanabe, H. Yanagi, T. Kamiya, Y. Kamihara, H. Hiramatsu, M. Hirano, and H. Hosono, Nickel-Based Oxyphosphide Superconductor with a Layered Crystal Structure, LaNiOP, Inorg. Chem. **46**, 7719 (2007).

[79] M. K. Wu, J. R. Ashburn, C. J. Torng, P. H. Hor, R. L. Meng, L. Gao, Z. J. Huang, Y. Q. Wang, and C. W. Chu, Superconductivity at 93 K in a new mixed-phase Y-Ba-Cu-O compound system at ambient pressure, Phys. Rev. Lett. **58**, 908 (1987).

[80] J. L. Tallon, R. G. Buckley, P. W. Gilberd, M. R. Presland, I. W. M. Brown, M. E. Bowden, L. A. Christian, and R. Goguel, High-Tc superconducting phases in the series Bi2.1(Ca, Sr)n+lCunO2n+4+δ, Nature **333**, 153 (1988).

[81] H. Maeda, Y. Tanaka, M. Fukutomi, and T. Asano, A New High-$T_c$ Oxide Superconductor without a Rare Earth Element, Jpn. J. Appl. Phys. **27**, L209 (1988).

[82] E. K. Ko et al., Signatures of ambient pressure superconductivity in thin film La3Ni2O7, Nature **638**, 935 (2025).

[83] Y. Lee et al., Synthesis of superconducting freestanding infinite-layer nickelate heterostructures on the millimetre scale, Nature Synthesis (2025).

[84] Y.-Q. Liu et al., Superconductivity at 31 K in a Stage-1 Sodium-Intercalated Graphite, J. Am. Chem. Soc. (2026).

[85] S. Park et al., Superconductivity emerging from a stripe charge order in IrTe2 nanoflakes, Nat. Commun. **12**, 3157 (2021).

[86] Z. Z. Sheng and A. M. Hermann, Bulk superconductivity at 120 K in the Tl–Ca/Ba–Cu–O system, Nature **332**, 138 (1988).

[87] S. N. Putilin, E. V. Antipov, O. Chmaissem, and M. Marezio, Superconductivity at 94 K in HgBa2Cu04+δ, Nature **362**, 226 (1993).

[88] N. Wang et al., Bulk high-temperature superconductivity in pressurized tetragonal La2PrNi2O7, Nature **634**, 579 (2024).

[89] D. V. Semenok et al., Stability and Superconductivity of Ternary Polyhydrides, Ann. Phys. **538**, (2026).

[90] I. A. Troyan et al., Anomalous High-Temperature Superconductivity in $YH_6$, Advanced Materials **33**, (2021).

[91] E. F. Talantsev, N. G. Shelushinina, T. P. Krinitsina, and A. S. Moskvin, Ten Years of Near-Room-Temperature Superconductivity and the Profound Legacy of Mikhail Eremets, Physics of Metals and Metallography **126**, 1127 (2025).

[92] R. Matsumoto, S. Nakano, T. Irifune, and Y. Takano, Development of Boron-Doped Diamond Electrodes in Diamond Anvil Cell and Application for Exploration of Superconducting Hydrides, Ann. Phys. **538**, (2026).

[93] S. Cross et al., Synthesis and stability of high-$T_c$ LaH$_{10\pmδ}$ films at high pressures, ArXiv (2026).

[94] Y. Li et al., Pressure-induced evolution of superconductivity and topological electronic structure in LaNiGa2, Phys. Rev. B **113**, 104505 (2026).

[95] E. F. Talantsev, S. C. Wimbush, N. M. Strickland, J. A. Xia, P. D'Souza, J. G. Storey, J. L. Tallon, B. Ingham, R. Knibbe, and N. J. Long, Oxygen Deficiency, Stacking Faults and Calcium Substitution in MOD YBCO Coated Conductors, IEEE Transactions on Applied Superconductivity **23**, 7200205 (2013).

[96] E. F. Talantsev, N. M. Strickland, S. C. Wimbush, J. G. Storey, J. L. Tallon, and N. J. Long, Hole doping dependence of critical current density in YBa2Cu3O7−δ conductors, Appl. Phys. Lett. **104**, (2014).

[97] C. Yelpo, S. Favre, R. Faccio, and D. Ariosa, Thickness-dependent structural, vibrational, and superconducting properties of Bi2Sr2CaCu2O8+x thin films, Journal of Physics: Condensed Matter **38**, 015405 (2026).

[98] S. A. Kuzmichev, I. V. Morozov, A. I. Shilov, Ye. O. Rakhmanov, and T. E. Kuzmicheva, Doping Evolution of the Superconducting Order Parameter of Na(Fe, Co)As Alkali-Metal Pnictides, JETP Lett. **120**, 125 (2024).

[99] S. Licciardello, J. Buhot, J. Lu, J. Ayres, S. Kasahara, Y. Matsuda, T. Shibauchi, and N. E. Hussey, Electrical resistivity across a nematic quantum critical point, Nature **567**, 213 (2019).

[100] S. Cai et al., The breakdown of both strange metal and superconducting states at a pressure-induced quantum critical point in iron-pnictide superconductors, Nat. Commun. **14**, 3116 (2023).

[101] Y.-T. Hsu, K. Lee, S. Badoux, C. Duffy, A. Cuoghi, B. Y. Wang, A. Kool, I. Haïk-Dunn, H. Y. Hwang, and N. E. Hussey, Transport phase diagram and anomalous metallicity in superconducting infinite-layer nickelates, Nat. Commun. **15**, 9863 (2024).

[102] K. Lee et al., Linear-in-temperature resistivity for optimally superconducting (Nd,Sr)NiO2, Nature **619**, 288 (2023).

[103] K. Mydeen, E. Lengyel, Z. Deng, X. C. Wang, C. Q. Jin, and M. Nicklas, Temperature-pressure phase diagram of the superconducting iron pnictide LiFeP, Phys. Rev. B **82**, 014514 (2010).

[104] F. Rullier-Albenque, D. Colson, A. Forget, and H. Alloul, Multiorbital Effects on the Transport and the Superconducting Fluctuations in LiFeAs, Phys. Rev. Lett. **109**, 187005 (2012).

[105] M. Osada et al., Bridging ambient- and high-pressure superconductivity in La2LnNi2O7 films, Nat. Mater. (2026).

[106] J. P. Sun, Z.-C. Wang, Z. Y. Liu, S. X. Xu, T. Eto, Y. Sui, B. S. Wang, Y. Uwatoko, G.-H. Cao, and J.-G. Cheng, Effect of pressure on the self-hole-doped superconductor $RbGd_2Fe_4As_4O_2$, Journal of Physics: Condensed Matter **31**, 044001 (2019).

[107] Y. Rakhmanov, S. Gavrilkin, A. Levakhova, S. Kuzmichev, I. Morozov, A. Shilov, and T. Kuzmicheva, Effect of Na Doping on the Structural, Superconductive and Magnetic Properties of Iron-Based Chalcogenides (Na, K,Rb)xFe2−ySe2, J. Supercond. Nov. Magn. **39**, 24 (2026).

[108] E. F. Talantsev, The dominance of non-electron–phonon charge carrier interaction in highly-compressed superhydrides, Supercond. Sci. Technol. **34**, 115001 (2021).

[109] P. Ch. Sahu, M. Yousuf, and K. G. Rajan, Electrical resistivity and phase-transition behaviour of uranium under pressure and temperature, Physica B Condens. Matter **183**, 145 (1993).

[110] G. T. Meaden, *Electrical Resistance of Metals* (Springer US, Boston, MA, 1965).

[111] J. M. Fournier, Bulk properties of the actinides, Journal of the Less Common Metals **121**, 101 (1986).

[112] A. Gangwar, K. Yadav, S. Patnaik, and S. Chatterjee, Evidence for fully-gapped superconductivity in BCS superconductor NiBi3, ArXiv (2026).

[113] M. Dormeval, N. Baclet, C. Valot, P. Rofidal, and J. M. Fournier, Crystalline and electronic structure of Pu–Ce and Pu–Ce–Ga alloys stabilized in the δ phase, J. Alloys Compd. **350**, 86 (2003).

[114] M. Gurvitch, Ioffe-Regel criterion and resistivity of metals, Phys. Rev. B **24**, 7404 (1981).

[115] H. Wiesmann, M. Gurvitch, H. Lutz, A. Ghosh, B. Schwarz, M. Strongin, P. B. Allen, and J. W. Halley, Simple Model for Characterizing the Electrical Resistivity in A−15 Superconductors, Phys. Rev. Lett. **38**, 782 (1977).

[116] M. Gurvitch, *RESISTIVITY OF A15 COMPOUNDS*, in *Superconductivity in D- and F-Band Metals* (Elsevier, 1980), pp. 317–334.

[117] E. Talantsev, Quantifying the Charge Carrier Interaction in Metallic Twisted Bilayer Graphene Superlattices, Nanomaterials **11**, 1306 (2021).

[118] A. F. Ioffe and A. R. Regel, *Non-Crystalline, Amorphous and Liquid Electronic Semiconductors* (John Wiley & Sons, New York, 1960).

[119] E. F. Talantsev, Advanced McMillan's equation and its application for the analysis of highly-compressed superconductors, Supercond. Sci. Technol. **33**, 094009 (2020).

[120] E. F. Talantsev, The electron–phonon coupling constant and the Debye temperature in polyhydrides of thorium, hexadeuteride of yttrium, and metallic hydrogen phase III, J. Appl. Phys. **130**, 195901 (2021).

[121] Y. Imai, F. Nabeshima, T. Yoshinaka, K. Miyatani, R. Kondo, S. Komiya, I. Tsukada, and A. Maeda, Superconductivity at 5.4 K in β-$Bi_2Pd$, J. Physical Soc. Japan **81**, 113708 (2012).

[122] B. Lv et al., Type-I superconductivity in noncentrosymmetric NbGe2, Phys. Rev. B **102**, 064507 (2020).

[123] V. K. Anand, A. D. Hillier, D. T. Adroja, A. M. Strydom, H. Michor, K. A. McEwen, and B. D. Rainford, Specific heat and μSR study on the noncentrosymmetric superconductor LaRhSi3, Phys. Rev. B **83**, 064522 (2011).

[124] E. F. Talantsev and E. G. Valova-Zaharevskaya, Structural and Superconducting Parameters of Highly Compressed Sulfur, Ann. Phys. **537**, e00084 (2025).

[125] E. F. Talantsev, Einstein, Debye, and Fermi Temperatures, Electron-Phonon Coupling Constant and Parabolic Pressure Dependence of the Transition Temperature in the $Y_{0.5}Ce_{0.5}H_9$ Hydride Superconductor, Ann. Phys. **538**, (2026).

[126] E. F. Talantsev and V. V Chistyakov, The A-15-type superconducting hydride La4H23: a nanograined structure with low strain, strong electron-phonon interaction, and a moderate level of nonadiabaticity, Supercond. Sci. Technol. **37**, 095016 (2024).

[127] W. Nowak, B. Rusin, M. Babij, R. Topolnicki, T. Ossowski, A. Pikul, and R. Idczak, Superconductivity in a New High-Entropy Alloy (NbTi)$$_{0.67}$$(MoHfV)$$_{0.33}$$, Metallurgical and Materials Transactions A **55**, 3789 (2024).

[128] D. A. Mayoh, J. A. T. Barker, R. P. Singh, G. Balakrishnan, D. McK. Paul, and M. R. Lees, Superconducting and normal-state properties of the noncentrosymmetric superconductor Re6Zr, Phys. Rev. B **96**, 064521 (2017).

[129] D. A. Mayoh, S. J. R. Holt, T. Takabatake, G. Balakrishnan, and M. R. Lees, Anisotropic superconductivity and unusually robust electronic critical field in single crystal La7Ir3, Phys. Rev. Mater. **5**, 114803 (2021).

[130] T. M. McQueen et al., Extreme sensitivity of superconductivity to stoichiometry in Fe1+δSe, Phys. Rev. B **79**, 014522 (2009).

[131] J.-Q. Yan et al., Magnetic and structural transitions in La0.4Na0.6Fe2As2 single crystals, Phys. Rev. B **91**, 024501 (2015).

[132] R. Kappenberger, S. Aswartham, F. Scaravaggi, C. G. F. Blum, M. I. Sturza, A. U. B. Wolter, S. Wurmehl, and B. Büchner, Solid state single crystal growth of three-dimensional faceted LaFeAsO crystals, J. Cryst. Growth **483**, 9 (2018).

[133] E. F. Talantsev, Quantifying interaction mechanism in infinite layer nickelate superconductors, J. Appl. Phys. **134**, (2023).

[134] E. F. Talantsev and V. V. Chistyakov, Debye temperature, electron-phonon coupling constant, and three-dome shape of crystalline strain as a function of pressure in highly compressed La3Ni2O7-δ, Letters on Materials **14**, 262 (2024).

[135] A. Pisoni, S. Katrych, P. Szirmai, B. Náfrádi, R. Gaál, J. Karpinski, and L. Forró, Upper critical field, pressure-dependent superconductivity and electronic anisotropy of $Sm_4Fe_2As_2Te_{1-x}O_{4-y}F_y$, Journal of Physics: Condensed Matter **28**, 115701 (2016).

[136] Y. Liu et al., Superconductivity and normal-state transport in compressively strained La2PrNi2O7 thin films, Nat. Mater. **24**, 1221 (2025).

[137] R. A. Cooper et al., Anomalous Criticality in the Electrical Resistivity of $La_{2-x}Sr_xCuO_4$, Science (1979). **323**, 603 (2009).

[138] D. P. Shoemaker, D. Y. Chung, H. Claus, M. C. Francisco, S. Avci, A. Llobet, and M. G. Kanatzidis, Phase relations in KxFe2-ySe2 and the structure of superconducting KxFe2Se2 via high-resolution synchrotron diffraction, Phys. Rev. B **86**, 184511 (2012).

[139] T. Kuzmicheva, L. Morgun, S. Gavrilkin, S. Kuzmichev, A. Degtyarenko, A. Muratov, I. Zhuvagin, A. Shilov, Y. Rakhmanov, and I. Morozov, Single Crystal Growth, Transport Phenomena, and Upper Critical Field of Alkali Metal-Based KxFe{2-y}(Se,S)2 Iron Chalcogenides, J. Supercond. Nov. Magn. **38**, 120 (2025).

[140] M. Roslova, S. Kuzmichev, T. Kuzmicheva, Y. Ovchenkov, M. Liu, I. Morozov, A. Boltalin, A. Shevelkov, D. Chareev, and A. Vasiliev, Crystal growth, transport phenomena and two-gap superconductivity in the mixed alkali metal (K(1−z)Na(z))xFe(2−y)Se2 iron selenide, CrystEngComm **16**, 6919 (2014).

[141] E. F. Talantsev, Universal temperature-dependent electrical resistivity in actinides, ArXiv (2026).

[142] N. F. Mott, Electrons in transition metals, Adv. Phys. **13**, 325 (1964).

[143] A. L. Efros and B. I. Shklovskii, Coulomb gap and low temperature conductivity of disordered systems, Journal of Physics C: Solid State Physics **8**, L49 (1975).

[144] N. Martin, H. Gerami, A. J. Santos, F. M. Morales, F. Amiot, J.-M. Cote, J. Gavoille, J.-B. Sanchez, and A. Besnard, Deposition angle tunes low-temperature electronic transport in early transition metal thin films, Mater. Des. **266**, 116075 (2026).

[145] V. J. Emery and S. A. Kivelson, Importance of phase fluctuations in superconductors with small superfluid density, Nature **374**, 434 (1995).

[146] E. F. Talantsev, W. P. Crump, J. G. Storey, and J. L. Tallon, London penetration depth and thermal fluctuations in the sulphur hydride 203 K superconductor, Ann. Phys. **529**, 1 (2017).

[147] L. G. Aslamasov and A. I. Larkin, The influence of fluctuation pairing of electrons on the conductivity of normal metal, Phys. Lett. A **26**, 238 (1968).

[148] M. Mondal, Phase fluctuations in a conventional s-wave superconductor: Role of dimensionality and disorder, ArXiv (2013).

[149] E. F. Talantsev, Two-Band Superconductivity and Transition Temperature Limited by Thermal Fluctuations in Ruddlesden–Popper Nickelates at Ambient and High Pressures, Ann. Phys. **537**, (2025).

[150] J. L. Tallon, J. G. Storey, and J. W. Loram, Fluctuations and critical temperature reduction in cuprate superconductors, Phys. Rev. B **83**, 092502 (2011).

[151] S. R. Ghorbani and X. L. Wang, In-field Conductivity Fluctuations in Ba0.72K0.28Fe2As2 Single Crystals, J. Supercond. Nov. Magn. **31**, 2349 (2018).

[152] S. Mandal, S. Dutta, S. Basistha, I. Roy, J. Jesudasan, V. Bagwe, L. Benfatto, A. Thamizhavel, and P. Raychaudhuri, Destruction of superconductivity through phase fluctuations in ultrathin <math> <mi>a</mi> </math> -MoGe films, Phys. Rev. B **102**, 060501 (2020).

[153] Jasper van Baten, *ScanIt*, 2.0.8.0.

[154] E. Seifert, OriginPro 9.1: Scientific Data Analysis and Graphing Software—Software Review, J. Chem. Inf. Model. **54**, 1552 (2014).

[155] Y. Ma, K. Hu, Q. Ji, B. Gao, H. Zhang, G. Mu, F. Huang, and X. Xie, Growth and characterization of CaFe1-Co AsF single crystals by CaAs flux method, J. Cryst. Growth **451**, 161 (2016).

[156] N. D. Zhigadlo, R. Puzniak, P. J. W. Moll, F. Bernardini, and T. Shiroka, Emergence of superconductivity in single-crystalline LaFeAsO under simultaneous Sm and P substitution, J. Alloys Compd. **958**, 170384 (2023).

[157] P. F. S. Rosa et al., Possible unconventional superconductivity in substituted BaFe2As2 revealed by magnetic pair-breaking studies, Sci. Rep. **4**, 6252 (2014).

[158] T. M. Garitezi et al., Synthesis and Characterization of BaFe2As2 Single Crystals Grown by In-flux Technique, Brazilian Journal of Physics **43**, 223 (2013).

[159] H. Q. Luo, P. Cheng, Z. S. Wang, H. Yang, Y. Jia, L. Fang, C. Ren, L. Shan, and H. H. Wen, Normal state transport properties in single crystals of Ba1−xKxFe2As2 andNdFeAsO1−xFx, Physica C Supercond. **469**, 477 (2009).

[160] G. F. Chen, Z. Li, J. Dong, G. Li, W. Z. Hu, X. D. Zhang, X. H. Song, P. Zheng, N. L. Wang, and J. L. Luo, Transport and anisotropy in single-crystalline SrFe2As2 and A0.6K0.4Fe2As2 (A=Sr, Ba) superconductors, Phys. Rev. B **78**, 224512 (2008).

[161] R. Hu, K. Cho, H. Kim, H. Hodovanets, W. E. Straszheim, M. A. Tanatar, R. Prozorov, S. L. Bud'ko, and P. C. Canfield, Anisotropic magnetism, resistivity, London penetration depth and magneto-optical imaging of superconducting $K_{0.80}$ $Fe_{1.76}$ $Se_{2}$ single crystals, Supercond. Sci. Technol. **24**, 065006 (2011).

[162] J. H. Tapp, Z. Tang, B. Lv, K. Sasmal, B. Lorenz, P. C. W. Chu, and A. M. Guloy, LiFeAs: An intrinsic FeAs-based superconductor with Tc<18 K, Phys. Rev. B **78**, 060505 (2008).

[163] D. R. Parker, M. J. Pitcher, P. J. Baker, I. Franke, T. Lancaster, S. J. Blundell, and S. J. Clarke, Structure, antiferromagnetism and superconductivity of the layered iron arsenide NaFeAs, Chemical Communications 2189 (2009).

[164] S. Kasahara, K. Hashimoto, H. Ikeda, T. Terashima, Y. Matsuda, and T. Shibauchi, Contrasts in electron correlations and inelastic scattering between LiFeP and LiFeAs revealed by charge transport, Phys. Rev. B **85**, 060503 (2012).

[165] O. Heyer, T. Lorenz, V. B. Zabolotnyy, D. V. Evtushinsky, S. V. Borisenko, I. Morozov, L. Harnagea, S. Wurmehl, C. Hess, and B. Büchner, Resistivity and Hall effect of LiFeAs: Evidence for electron-electron scattering, Phys. Rev. B **84**, 064512 (2011).

[166] F. Rullier-Albenque, D. Colson, A. Forget, and H. Alloul, Erratum: Multiorbital Effects on the Transport and the Superconducting Fluctuations in LiFeAs [Phys. Rev. Lett. **109** , 187005 (2012)], Phys. Rev. Lett. **113**, 209901 (2014).

[167] Y. Liu, Y.-B. Liu, Q. Chen, Z.-T. Tang, W.-H. Jiao, Q. Tao, Z.-A. Xu, and G.-H. Cao, A new ferromagnetic superconductor: CsEuFe4As4, Sci. Bull. (Beijing). **61**, 1213 (2016).

[168] J.-K. Bao, K. Willa, M. P. Smylie, H. Chen, U. Welp, D. Y. Chung, and M. G. Kanatzidis, Single Crystal Growth and Study of the Ferromagnetic Superconductor $RbEuFe_4As_4$, Cryst. Growth Des. **18**, 3517 (2018).

[169] M. Bristow, W. Knafo, P. Reiss, W. Meier, P. C. Canfield, S. J. Blundell, and A. I. Coldea, Competing pairing interactions responsible for the large upper critical field in a stoichiometric iron-based superconductor CaKFe4As4, Phys. Rev. B **101**, 134502 (2020).

[170] S. Pyon et al., Large and significantly anisotropic critical current density induced by planar defects in CaKFe4As4 single crystals, Phys. Rev. B **99**, 104506 (2019).

[171] S. Katrych, K. Rogacki, A. Pisoni, S. Bosma, S. Weyeneth, R. Gaal, N. D. Zhigadlo, J. Karpinski, and L. Forró, Pr4Fe2As2Te1−xO4: A layered FeAs-based superconductor, Phys. Rev. B **87**, 180508 (2013).

[172] S. Medvedev et al., Electronic and magnetic phase diagram of β-Fe1.01Se with superconductivity at 36.7 K under pressure, Nat. Mater. **8**, 630 (2009).

[173] W.-H. Zhang et al., Direct Observation of High-Temperature Superconductivity in One-Unit-Cell FeSe Films, Chinese Physics Letters **31**, 017401 (2014).

[174] Q.-Y. Wang et al., Interface-Induced High-Temperature Superconductivity in Single Unit-Cell FeSe Films on $SrTiO_3$, Chinese Physics Letters **29**, 037402 (2012).

[175] J.-F. Ge, Z.-L. Liu, C. Liu, C.-L. Gao, D. Qian, Q.-K. Xue, Y. Liu, and J.-F. Jia, Superconductivity above 100 K in single-layer FeSe films on doped SrTiO3, Nat. Mater. **14**, 285 (2015).

[176] E. F. Talantsev, W. P. Crump, J. O. Island, Y. Xing, Y. Sun, J. Wang, and J. L. Tallon, On the origin of critical temperature enhancement in atomically thin superconductors, 2d Mater. **4**, 025072 (2017).

[177] I. Pallecchi, G. Lamura, M. Tropeano, M. Putti, R. Viennois, E. Giannini, and D. Van der Marel, Seebeck effect in Fe1+xTe1−ySey single crystals, Phys. Rev. B **80**, 214511 (2009).

[178] T. J. Liu et al., Charge-carrier localization induced by excess Fe in the superconductor Fe1+yTe1−xSex, Phys. Rev. B **80**, 174509 (2009).

[179] Y. Sun, T. Taen, T. Yamada, Y. Tsuchiya, S. Pyon, and T. Tamegai, Evolution of superconducting and transport properties in annealed $FeTe_{1-x}Se_x (0.1 \leq x \leq 0.4)$ multiband superconductors, Supercond. Sci. Technol. **28**, 044002 (2015).

[180] Y. Sun, T. Taen, Y. Tsuchiya, Z. X. Shi, and T. Tamegai, Effects of annealing, acid and alcoholic beverages on $Fe_{1+y}Te_{0.6}Se_{0.4}$, Supercond. Sci. Technol. **26**, 015015 (2013).

[181] V. Tsurkan, J. Deisenhofer, A. Günther, Ch. Kant, M. Klemm, H.-A. Krug von Nidda, F. Schrettle, and A. Loidl, Physical properties of FeSe0.5Te0.5 single crystals grown under different conditions, Eur. Phys. J. B **79**, 289 (2011).

[182] V. Ksenofontov, G. Wortmann, A. I. Chumakov, T. Gasi, S. Medvedev, T. M. McQueen, R. J. Cava, and C. Felser, Density of phonon states in superconducting FeSe as a function of temperature and pressure, Phys. Rev. B **81**, 184510 (2010).

[183] Z. Li, Y. Li, Z. Wang, G. Cao, B. Zhang, H. Pang, and F. Li, Study of the Rare Earth Effects on the Magnetic Fluctuations in RbLn2Fe4As4O2 (Ln = $=$ Tb, Dy, and Ho) by Mössbauer Spectroscopy, J. Supercond. Nov. Magn. **32**, 361 (2019).

[184] M. A. Albedah, F. Nejadsattari, Z. M. Stadnik, Y. Liu, and G.-H. Cao, Mössbauer spectroscopy measurements on the 35.5 K superconductor Rb1−δEuFe4As4, Phys. Rev. B **97**, 144426 (2018).

[185] W. L. McMillan, Transition Temperature of Strong-Coupled Superconductors, Physical Review **167**, 331 (1968).

[186] R. C. Dynes, McMillan's equation and the Tc of superconductors, Solid State Commun. **10**, 615 (1972).

[187] B. C. Sales, A. S. Sefat, M. A. McGuire, R. Y. Jin, D. Mandrus, and Y. Mozharivskyj, Bulk superconductivity at 14 K in single crystals of Fe1+yTexSe1−x, Phys. Rev. B **79**, 094521 (2009).

[188] U. Stockert, M. Abdel-Hafiez, D. V. Evtushinsky, V. B. Zabolotnyy, A. U. B. Wolter, S. Wurmehl, I. Morozov, R. Klingeler, S. V. Borisenko, and B. Büchner, Specific heat and angle-resolved photoemission spectroscopy study of the superconducting gaps in LiFeAs, Phys. Rev. B **83**, 224512 (2011).

[189] B. Lee, S. Khim, J. S. Kim, G. R. Stewart, and K. H. Kim, Single-crystal growth and superconducting properties of LiFeAs, EPL (Europhysics Letters) **91**, 67002 (2010).

[190] A. F. Wang, J. J. Lin, P. Cheng, G. J. Ye, F. Chen, J. Q. Ma, X. F. Lu, B. Lei, X. G. Luo, and X. H. Chen, Phase diagram and physical properties of NaFe1−xCuxAs single crystals, Phys. Rev. B **88**, 094516 (2013).

[191] G. Mu, H. Luo, Z. Wang, L. Shan, C. Ren, and H.-H. Wen, Low temperature specific heat of the hole-doped Ba0.6K0.4Fe2As2 single crystals, Phys. Rev. B **79**, 174501 (2009).

[192] M. Rotter, M. Tegel, I. Schellenberg, F. M. Schappacher, R. Pöttgen, J. Deisenhofer, A. Günther, F. Schrettle, A. Loidl, and D. Johrendt, Competition of magnetism and superconductivity in underdoped (Ba (1-x)K(x))Fe2As2, New J. Phys. **11**, 025014 (2009).

[193] J. K. Dong, L. Ding, H. Wang, X. F. Wang, T. Wu, G. Wu, X. H. Chen, and S. Y. Li, Thermodynamic properties of $Ba_{1-x}K_xFe_2As_2$ and $Ca_{1-x}Na_xFe_2As_2$, New J. Phys. **10**, 123031 (2008).

[194] F. Ronning, T. Klimczuk, E. D. Bauer, H. Volz, and J. D. Thompson, Synthesis and properties of $CaFe_2As_2$ single crystals, Journal of Physics: Condensed Matter **20**, 322201 (2008).

[195] B. Zeng, B. Shen, G. F. Chen, J. B. He, D. M. Wang, C. H. Li, and H. H. Wen, Nodeless superconductivity of single-crystalline KFe2−ySe2 revealed by the low-temperature specific heat, Phys. Rev. B **83**, 144511 (2011).

[196] M. Roslova, S. Kuzmichev, T. Kuzmicheva, Y. Ovchenkov, M. Liu, I. Morozov, A. Boltalin, A. Shevelkov, D. Chareev, and A. Vasiliev, Crystal growth, transport phenomena and two-gap superconductivity in the mixed alkali metal (K{1−z}Na{z})xFe(2−y)Se2 iron selenide, CrystEngComm **16**, 6919 (2014).

[197] M. Tegel, S. Johansson, V. Weiß, I. Schellenberg, W. Hermes, R. Pöttgen, and D. Johrendt, Synthesis, crystal structure and spin-density-wave anomaly of the iron arsenide-fluoride SrFeAsF, EPL (Europhysics Letters) **84**, 67007 (2008).

[198] P. Singh, K. Kwatek, T. Zajarniuk, T. Palasyuk, C. Jastrzębski, A. Szewczyk, and S. J. Singh, Electronic phase diagram and enhanced superconductivity in fluorine-doped PrFeAsO1-x$F_x$, J. Alloys Compd. **1072**, 188942 (2026).

[199] M. Gang, Z. Xi-Yu, F. Lei, S. Lei, R. Cong, and W. Hai-Hu, Nodal Gap in Fe-Based Layered Superconductor $LaO_{0.9}F_{0.1-\delta}FeAs$ Probed by Specific Heat Measurements, Chinese Physics Letters **25**, 2221 (2008).

[200] Y. Xiao et al., Neutron diffraction study of phase transitions and thermal expansion of SrFeAsF, Phys. Rev. B **81**, 094523 (2010).

[201] S. A. J. Kimber et al., Magnetic ordering and negative thermal expansion in PrFeAsO, Phys. Rev. B **78**, 140503 (2008).